%% file: mbhbias.astroph.tex
\documentclass[useAMS,usenatbib]{mn2e}

\input{epsf}

\usepackage{graphicx}	
\usepackage{deluxetable}
\usepackage{amsmath}

\input{commands.tex}

\title[Orientation dependence of quasar black hole masses]{The orientation dependence of quasar single-epoch black hole mass scaling relationships}
\author[J. C. Runnoe et al.]{Jessie C. Runnoe$^{1}$\thanks{E-mail:
jrunnoe@uwyo.edu} , M. S. Brotherton$^{1}$, Z. Shang$^{2}$, B. J. Wills$^{3}$, and M. A. DiPompeo$^1$\\
$^{1}$Department of Physics and Astronomy, University of Wyoming, Laramie, WY 82071, USA\\
$^{2}$Department of Physics, Tianjin Normal University, Tianjin 300074, China \\
$^{3}$Department of Astronomy, University of Texas, Austin, TX 78712, USA}
\begin{document}		

\date{Preprint 2012 July 05}

\pagerange{\pageref{firstpage}--\pageref{lastpage}} \pubyear{2012}

\maketitle

\label{firstpage}

\begin{abstract}
Black hole masses are estimated for radio-loud quasars using several self-consistent scaling relationships based on emission-line widths and continuum luminosities.  The emission lines used, \Hb, \MgIIw, and \CIVw, have different dependencies on orientation as estimated by radio core dominance.  We compare differences in the log of black hole masses estimated from different emission lines and show that they depend on radio core dominance in the sense that core-dominated, jet-on objects have systematically smaller \Hb\ and \MgII\ determined masses compared to those from \CIV, while lobe-dominated edge-on objects have systematically larger \Hb\ and \MgII\ determined masses compared to those from \CIV.  The effect is consistent with the \Hb\ line width, and to a lesser extent that of \MgII, being dependent upon orientation in the sense of a axisymmetric velocity field plus a projection effect.  The size of the effect is nearly an order of magnitude in black hole mass going from one extreme orientation to the other.  We find that radio spectral index is a good proxy for radio core dominance and repeating this analysis with radio spectral index yields similar results.  Accounting for orientation could in principle significantly reduce the scatter in black hole mass scaling relationships, and we quantify and offer a correction for this effect cast in terms of radio core dominance and radio spectral index.
\end{abstract}


%INTRODUCTION
%%%%%%%%%%%%%%%%%%%%%%%%%%%%%%%%%%%%%%%%%%%%%%%%%%%%%%%%%%%%%%%%%%%%%%%%%%%%%%%%%
\section{introduction}	

Black hole mass is an important fundamental quasar parameter that can be measured directly via reverberation mapping, which probes the motions of gas in the vicinity of the black hole \citep{blandford82,peterson93}.  If the motion of gas in the broad line region (BLR) is dominated by the gravitational potential of the black hole, then the central mass can be estimated from the scaled virial product,   

\begin{equation}
M_{BH} = f\frac{\Delta V^2 R}{G},
\end{equation} 	

where $G$ is the gravitational constant, $R$ is the distance to the broad line region gas that has velocity dispersion $\Delta V$, and $f$ is a scale factor of order unity that contains unknown information about the BLR geometry and orientation effects.  

The factor $f$ may vary between objects and possibly even between emission lines in the same object, but determining $f$ for individual objects requires an independent measure of the black hole mass.  Instead, a measure of the statistical scale factor, $<f>$, is borrowed from the $M_{BH}-\sigma_{*}$ relationship for quiescent galaxies in order to calibrate the zero-point of the reverberation mapping masses.  One recent measurement of the statistical scale factor yields $<f>=5.25\pm1.21$ \citep{woo10}.  

Reverberation mapping then measures the central black hole mass by combining emission-line width ($\Delta V$) and the size of the BLR ($R_{BLR}$), inferred from the time delay in an emission line ($R_{BLR} = c \,\tau$).  At present, reverberation mapping black hole masses have been measured for approximately 50 active galactic nuclei (AGN) \citep[e.g.,][]{peterson04, bentz09c, grier12}.   

Another important result has followed from reverberation mapping: the radius-luminosity ($R-L$) relationship.  The time lag for emission lines to respond to variation in the continuum is longer in more luminous objects \citep{kaspi00,kaspi05}, such that $R_{BLR}\propto L^{1/2}$ when host galaxy contamination is removed \citep{bentz09}.  This result is based on measurements of \Hb, but limited \CIV-based results are consistent with this relationship \citep{kaspi07}.

The $R-L$ relationship anchors the single-epoch ``mass scaling relationships," which provide an indirect estimate of black hole mass.  The mass scaling relationships are widely used because they can be applied to any object with a single-epoch optical/ultra-violet (UV) spectrum measurement of a broad emission-line width and continuum luminosity, the proxy for the BLR radius.  The velocity dispersion is usually measured, for \Hb, \MgIIw, or \CIVw, either from the full width at half maximum (FWHM) or the second moment of the line, $\sigma_{line}$, called the line dispersion.  The two line measurements have different characteristics and advantages, though $\sigma_{line}$ may be preferred \citep{collin06,peterson11}.  We focus on the FWHM line measurement in order to facilitate comparisons with older studies and save an investigation of $\sigma_{line}$ for future work.

Multiple studies have shown that the \Hb\ FWHM measurements used to estimate black hole mass depend on orientation.  Radio core dominance ($R$), the ratio of the radio core flux density to the extended radio lobe flux density at 5 GHz rest-frame, is related to the angle between the radio axis and the line of sight in the beaming model for radio sources \citep[e.g.,][]{scheuer79,orr82,ghisellini93}.  \citet{wills86} found that the FWHM of \Hb\ was significantly correlated with radio core dominance such that there is an absence of very broad widths for objects with large core dominance.  Their result is consistent with the motion of the \Hb-emitting gas being primarily in a plane perpendicular to the radio axis.  \citet{mclure01} test such a disc model for the BLR and find that it is a good fit to observations for \Hb.

The relationship between broad-line FWHM and core dominance found for \Hb\ has also been investigated for other broad lines, sometimes using radio spectral index as a proxy for radio core dominance. Radio spectral index and radio core dominance are well correlated \citep{brotherton96,jarvis06} but it should be noted that radio spectral index is neither the same as radio core dominance nor a perfect measure of orientation \citep[e.g.,][]{dipompeo12}.  \citet{jarvis06} and \citet{fine11} find that the FWHM of \MgII\ has an orientation dependence weaker than that of \Hb\ and \citet{fine11} find no orientation dependence for the FWHM of \CIV, where radio spectral index is used to indicate orientation.  This result for \CIV\ is consistent with the result from \citet{vestergaard00} that FWHM \CIV\ does not correlate with radio core dominance.

Given the orientation dependence of FWHM for the broad emission lines used to estimate black hole mass, it has been suggested that black hole estimates based on single-epoch scaling relationships will depend on orientation \citep[e.g.,][]{baskin05,jarvis06}.  Black hole masses cannot be as accurately estimated without $\Delta$V \citep{assef12}, so we investigate the dependence of black hole mass estimates on radio core dominance for the radio-loud (RL) subsample from the \citet{shang11} spectral energy distribution (SED) atlas.

In Section~\ref{sec:data} we provide sample data and discuss spectral fitting and black hole mass estimates.  Section~\ref{sec:analysis} confirms the correlations between FWHM and radio core dominance and presents the correlations between differences in the log of black hole mass estimates and radio core dominance.  Here we also provide corrections to black hole mass  based on the radio core dominance.  To conclude Section~\ref{sec:analysis}, the analysis is repeated for radio spectral index.  The correlations between differences in the log of black hole mass estimated from different emission lines and radio core dominance are discussed in Section~\ref{sec:discussion} and Section~\ref{sec:conclusion} summarizes our investigation and corrections to black hole mass.

Black hole masses and luminosities are calculated using a cosmology with $H_0 = 70$ km s$^{-1}$ Mpc$^{-1}$, $\Omega_{\Lambda} = 0.7$, and $\Omega_{m} = 0.3$.

Because we employ several values of ``$R$'' in this study, we explicitly state the definition of each parameter here to avoid confusion.  Radio core dominance, $R$, is the ratio of the core flux density to the lobe flux density at 5~GHz rest frame.  The ratio of the core luminosity at 5~GHz rest frame to the optical V-band luminosity, $R_{\textrm{V}}$, is another measure of core dominance \citep{willsbro95}.  Radio loudness, $R^{*}$, is the ratio of the total radio luminosity at 5~GHz rest frame to the optical luminosity \citep{stocke92}.  The radio spectral index, $\alpha$ ($f_{\nu} \sim \nu^{\alpha}$), is a slope between two or more radio frequencies.

%DATA
%%%%%%%%%%%%%%%%%%%%%%%%%%%%%%%%%%%%%%%%%%%%%%%%%%%%%%%%%%%%%%%%%%%%%%%%%%%%%%%%%
\section{Sample, Data, and Measurements}
\label{sec:data}
For this investigation, we the RL subsample of the \citet{shang11} SED atlas.  See \citet{wills95} or \citet{netzer95} for additional sample details.   Many of the blazars originally included in this sample have been excluded from this investigation.  4C 11.69, which was identified as a blazar by \citet{runnoe12b} based on infrared variability, has also been excluded.  Our final sample has 52 objects.

\begin{figure*}
\begin{minipage}[!b]{8cm}
\centering
\includegraphics[width=8cm]{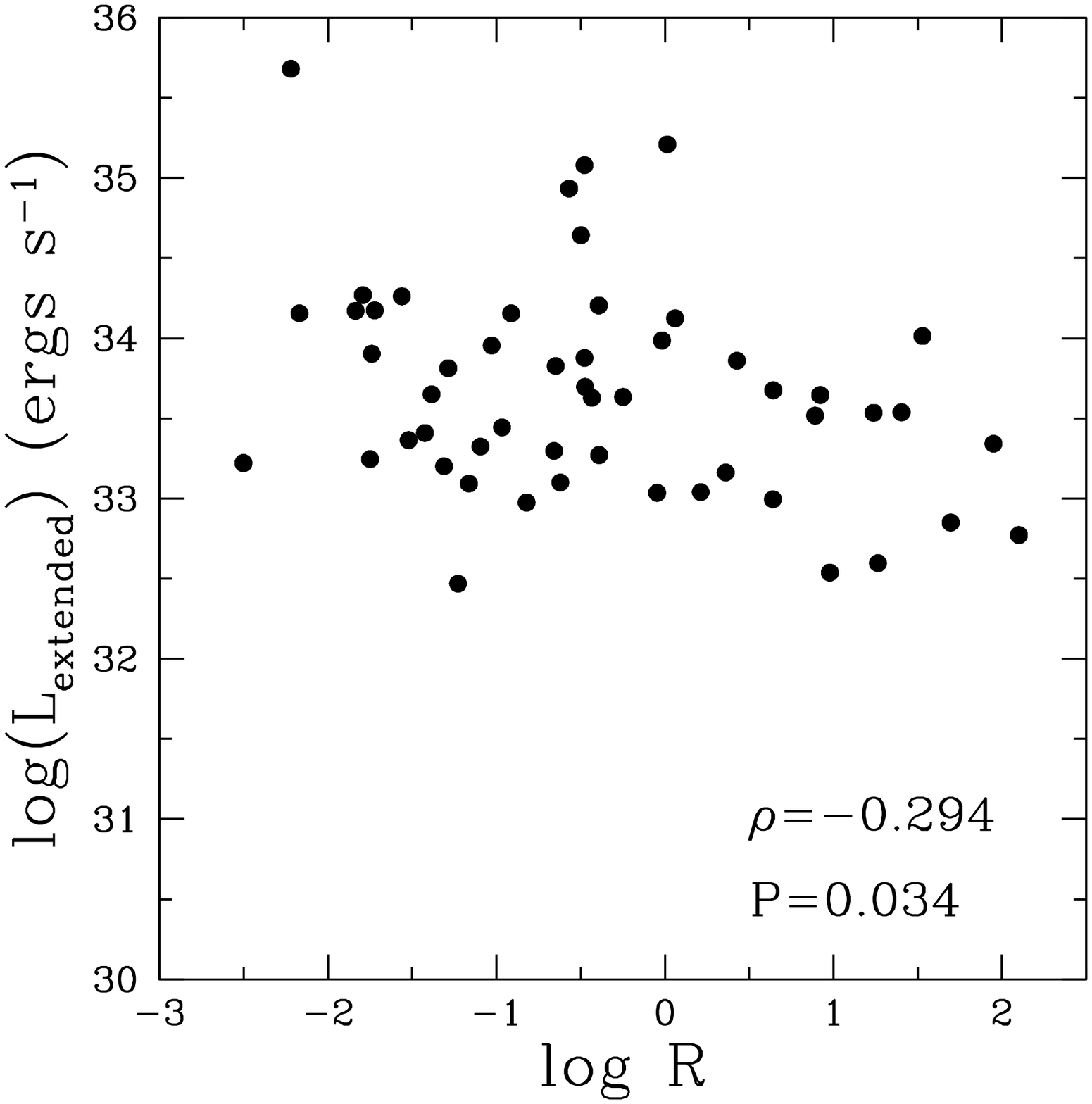}
\end{minipage}\hspace{0.6cm}
\hspace{0.6cm}
\begin{minipage}[!b]{8cm}
\centering
\includegraphics[width=8cm]{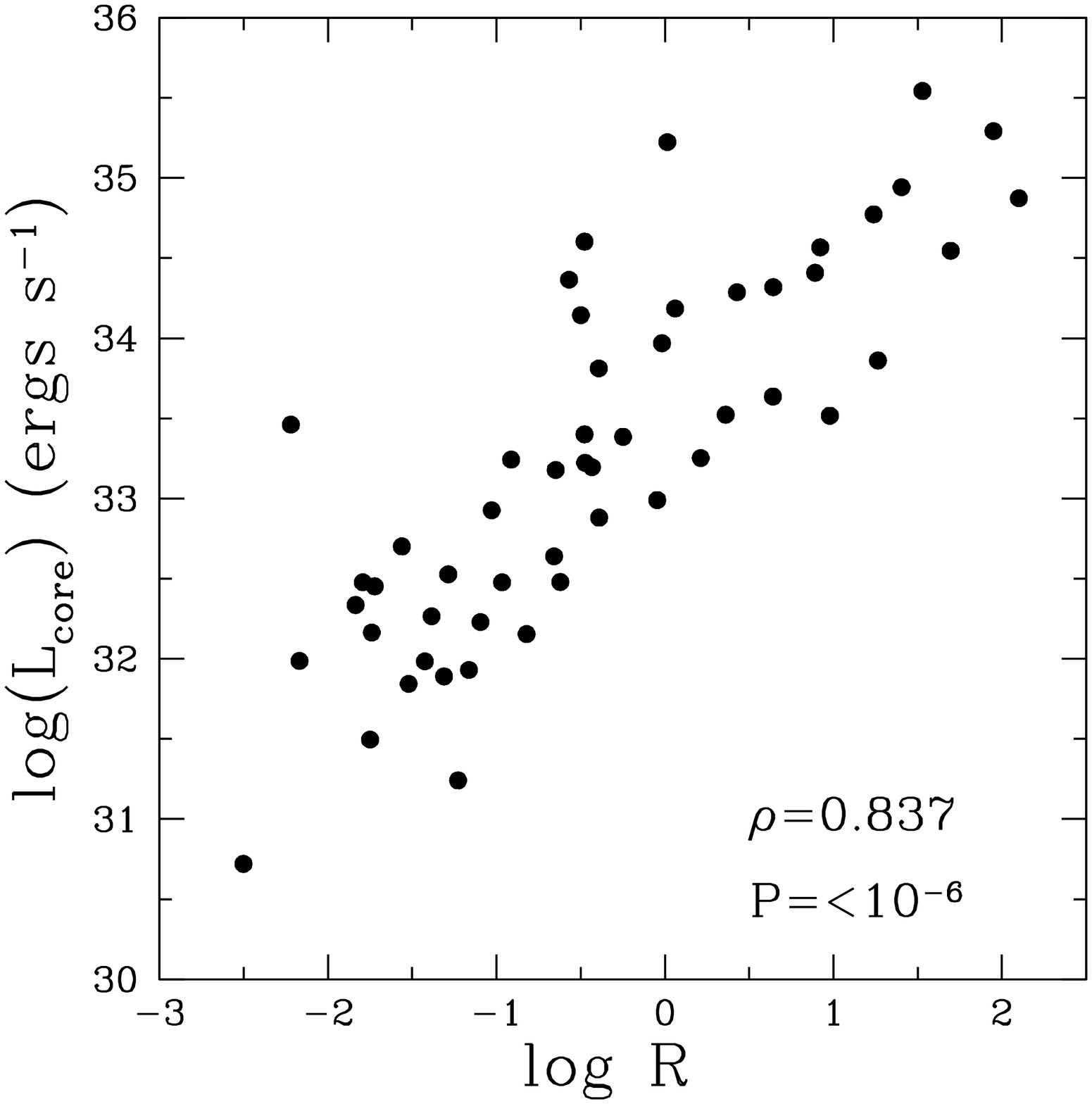}
\end{minipage}
\hspace{0.6cm}                   
\caption{Core and extended radio luminosities versus log~$R$.  The success of selecting objects with similar extended luminosities is apparent in the left panel, so variation in log~$R$ is due primarily to variation in the radio core luminosity seen in the right panel. \label{fig:selection}}
\end{figure*}

Originally assembled to study orientation, sample members were selected to have similar extended radio luminosity, which is thought to be isotropic.  Thus, objects in the sample are similar to each other but viewed at different angles indicated by their radio core luminosities.  While this selection was compromised by the original spherical aberration problem with Hubble Space Telescope, requiring selection of brighter targets, the extended radio luminosities are similar with a limited range versus log~$R$ (Figure~\ref{fig:selection}, left), and variation in radio core luminosity is therefore largely responsible for variation in log~$R$ (right).

We calculate radio core dominance, the ratio of the flat-spectrum core to the steep-spectrum extended lobes at 5~GHz rest frame, for the sample.  The radio data from the \citet{shang11} SEDs are insufficient for this calculation, so we assembled radio core and total flux densities from high quality surveys and the literature.  We applied a k-correction to the radio fluxes assuming slopes of $\alpha=0$ and $\alpha=-0.7$ for the core and lobes, respectively.  We also calculate the orientation indicator log~$R_{\textrm{V}}$, the 5 GHz core flux density to the optical V-band flux density.   

We calculate radio spectral index for our sample in order to provide an orientation measure that may be easier to obtain and can be measured in some cases when log~$R$ cannot.  We use the convention $f_{\nu} \sim \nu^{\alpha}$, where $f_{\nu}$ is the radio flux, $\nu$ is the frequency, and $\alpha$ is the radio spectral index.  We measure the slope in two ways from the radio data in the \citet{shang11} SEDs.  $\alpha_{all}$ is the radio spectral index measured by fitting a line to all of the available radio data in logarithmic space.  $\alpha^{1.4}_{4.9}$ is the radio spectral index measured from two data points at observed-frame 1.4 GHz and 4.9 GHz.  We include this second measure in order to avoid the turnover of the radio spectrum around $1$ GHz that is due to synchrotron self absorption and does not reflect the orientation of the object.  We do note that we do not observe such a turnover in most of our objects; most are well fit with a single line using all of the available radio data from the SEDs.

Table~\ref{tab:radio} lists radio properties for the sample.  It is arranged as follows:
\begin{itemize}
\item[] Column (1) gives the object name.
\item[] Column (2) gives the redshift from \citet{shang11}.
\item[] Column (3) gives the frequency at which fluxes were observed for the core and total radio components, respectively.  In cases where the extended radio flux was observed instead of the total, the frequency of this observation is listed in Column (3).
\item[] Columns (4) and (5) give the k-corrected, 5~GHz rest-frame flux densities in mJy.  In most cases, the extended flux is calculated from measurements of the core and total flux, but in some cases the extended and core emission were the quantities directly listed in the reference.  The cases where extended emission is actually observed rather than calculated are indicated in Column 5.  
\item[] Column (6) gives the log of the k-corrected, rest-frame, radio core dominance, log~$R$.
\item[] Column (7) gives the radio loudness parameter, log~$R^*$, where \citet{shang11} calculates $R^*=f(5 \textrm{ GHz})/f(4215 \textnormal{ \AA})$ in the rest frame.
\item[] Column (8) gives the reference for the radio data used to calculate radio core dominance.  Radio spectral index is always calculated from the radio portion of the \citet{shang11} SEDs.
\end{itemize}

\input{radiotab.tex}

Spectral fitting to measure emission-line widths and continuum luminosities was carried out by \citet{tang12} following the prescription of \citet{shang07}.  \citeauthor{tang12} fit the optical/UV region of the SEDs using {\sc specfit} \citep{kriss94}.  They model the regions around \CIVw, \MgIIw, and \Hb\ each with a power-law continuum and a line profile of two Gaussian components.  A third Gaussian, that is not included in the total FWHM of the line profile, is added to the \Hb\ profile to fit the contribution to emission from the narrow-line region.  The \CIV\ line does not have a strong contribution from the narrow-line region \citep{wills93} and, while \MgII\ does sometimes have such a component, it is not apparent in our objects.  Optical and UV \FeII\ emission is modeled with templates \citep{bg92,vestergaard01} based on the Seyfert 1 galaxy, I Zw1, that are allowed to vary in amplitude and velocity width.  Continuum luminosities based on flux measurements from these fits are from \citet{runnoe12a}.  The resulting FWHM values for \Hb, \CIV, and \MgII, the bolometric luminosity, and other relevant data are given in Table~\ref{tab:mbh}

\citet{tang12} estimate black hole masses for the \citeauthor{shang11} atlas using single-epoch scaling relationships from \citet{vestergaard06} for \Hb\ and \CIV\ and from \citet{vestergaard09} for \MgII.  Multiple lines are used to statistically approach the true black hole mass and these scaling relationships were chosen specifically because they are self-consistent \citep{vestergaard11}, thus enabling meaningful comparisons among them.  The optical/UV region of our SEDs facilitates consistent estimates of black hole mass from different broad emission lines because they each have quasi-simultaneous data and coverage over a large enough wavelength range to include \Hb, \MgII, and \CIV.  Black hole masses are listed in Table~\ref{tab:mbh}.

Not all measurements are available for all objects.  There are 38 (73\%) objects with measurements of log~$R$, \fwhmhb, and \fwhmciv, 39 (75\%) objects with measurements of log~$R$, \fwhmhb, and \fwhmmgii, and 51 (98\%) objects with measurements of log~$R$, \fwhmciv, and \fwhmmgii.

\input{mbhtab.tex}

%ANALYSIS
%%%%%%%%%%%%%%%%%%%%%%%%%%%%%%%%%%%%%%%%%%%%%%%%%%%%%%%%%%%%%%%%%%%%%%%%%%%%%%%%%
\section{Analysis}
\label{sec:analysis}
We investigate the dependence of the difference between the log of the black hole masses estimated with different scaling relationships on radio core dominance.  We first demonstrate the relationship between FWHM and radio core dominance for \Hb, \MgII, and \CIV, and then proceed to characterize the dependence this creates in black hole masses derived using scaling relationships calibrated for each of these emission lines.  We then repeat the analysis for radio spectral index.

\subsection{The orientation dependence of broad-line widths}
\begin{figure*}
\begin{minipage}[!b]{5cm}
\centering
\includegraphics[width=6cm]{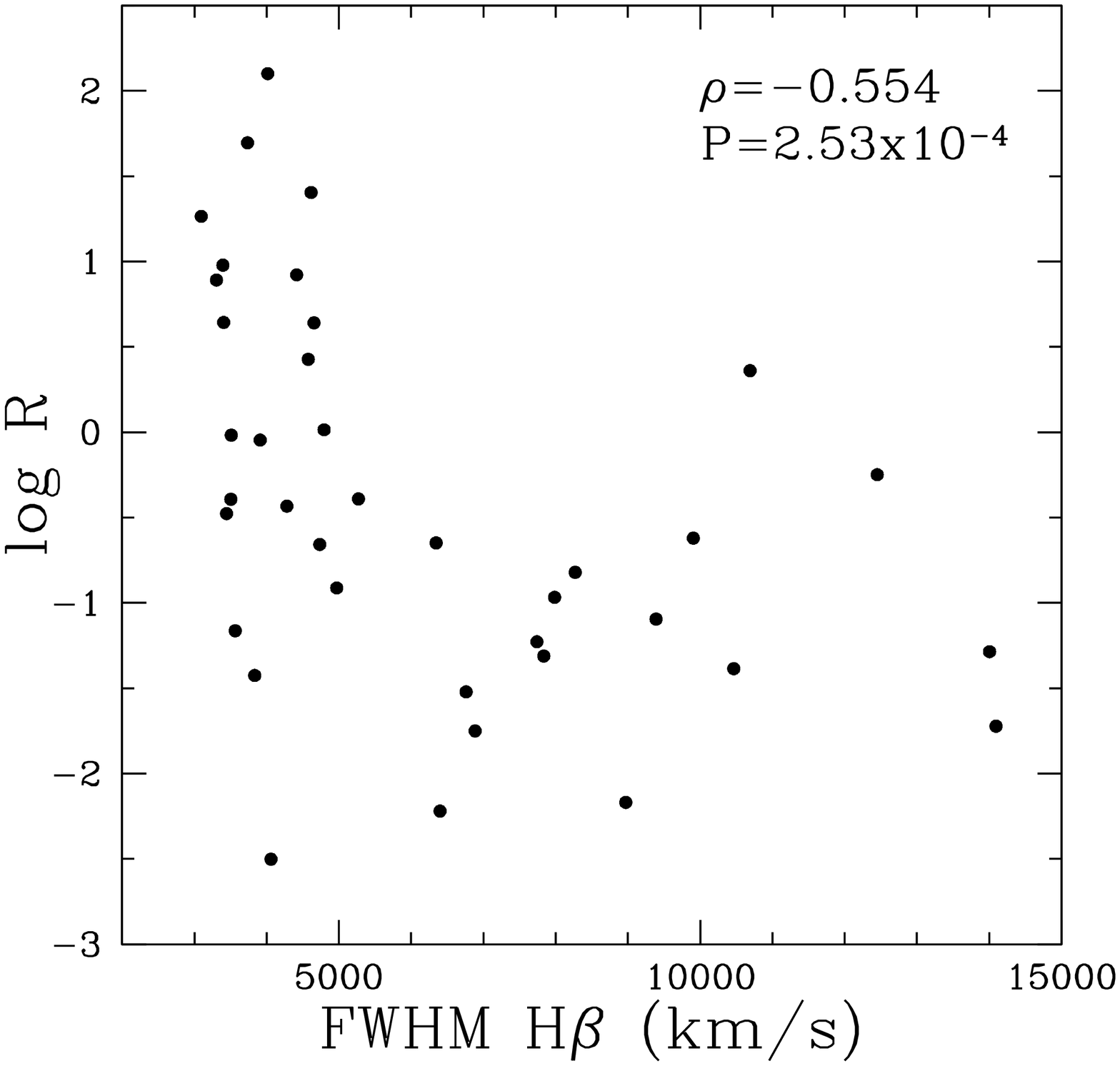}
\end{minipage}
\hspace{0.6cm}
\begin{minipage}[!b]{5cm}
\centering
\includegraphics[width=6cm]{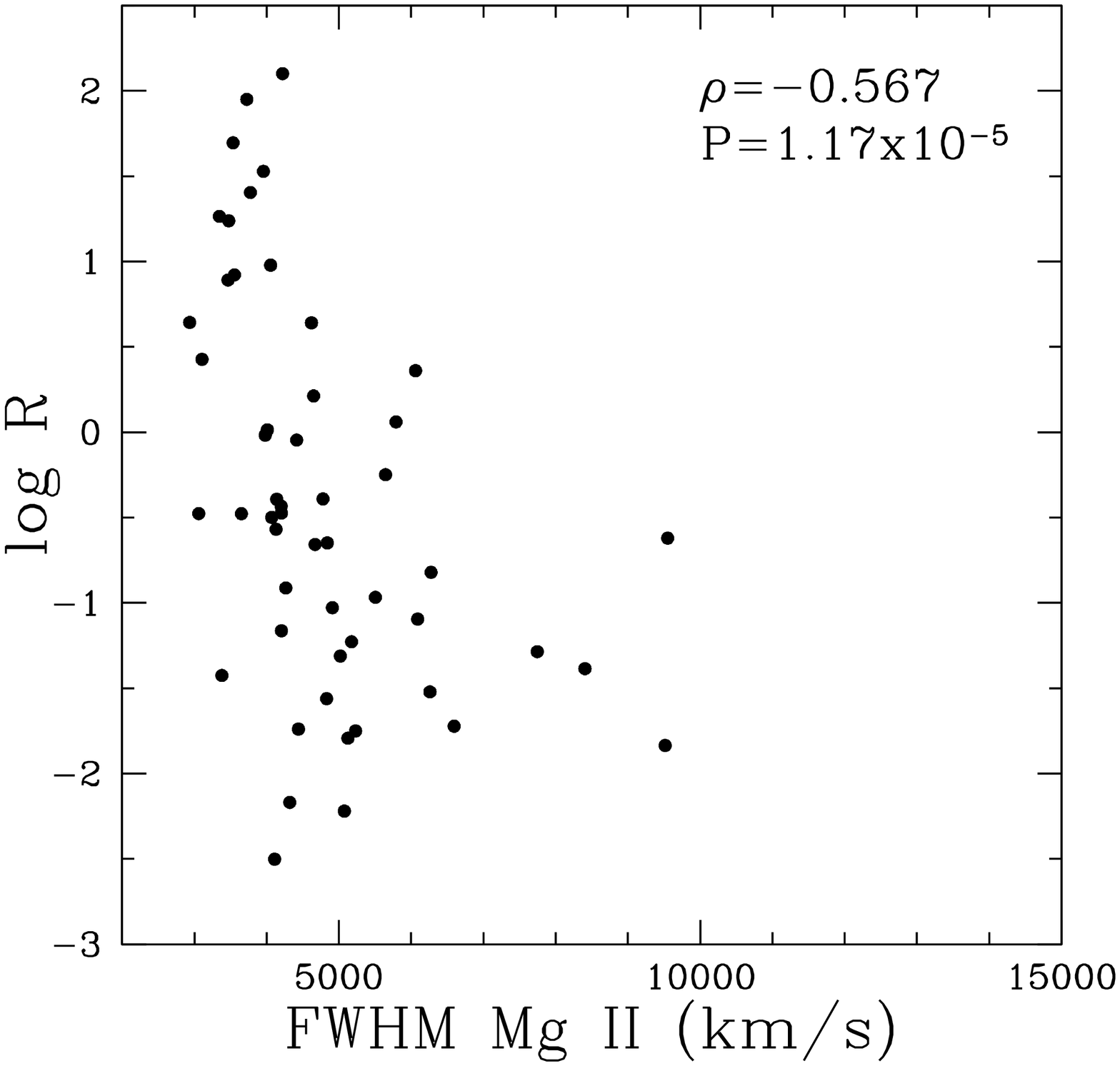}
\end{minipage}
\hspace{0.6cm}
\begin{minipage}[!b]{5cm}
\centering
\includegraphics[width=6cm]{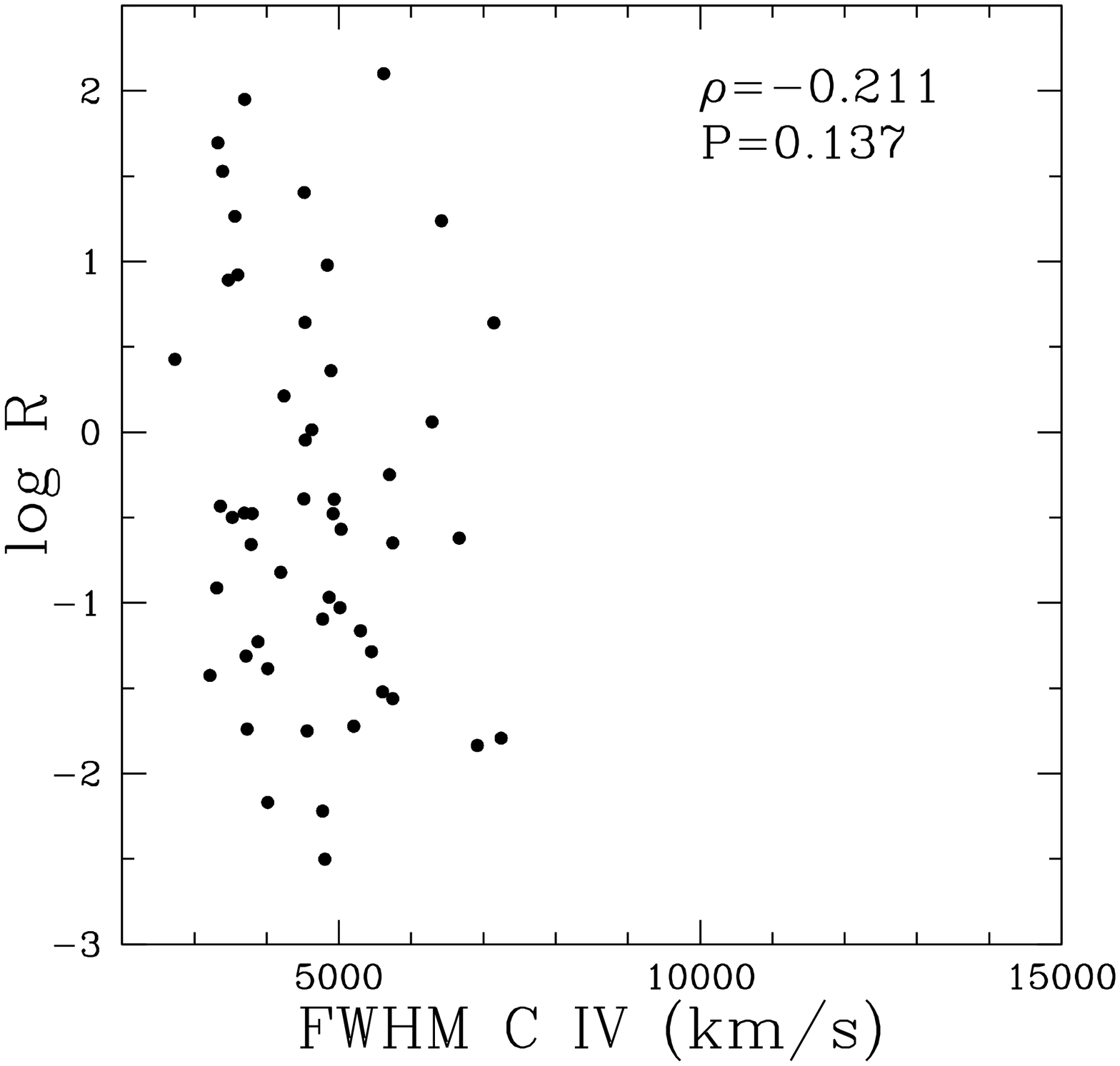}
\end{minipage}                             
\caption{Radio core dominance versus FWHM for \Hb, \MgII, and \CIV.  The effect seen in \citet{wills86}, where large values of FWHM have small vales of log~$R$ but small values of FWHM have a range of log~$R$, is reproduced for \Hb\ and, to some extent, \MgII.  This effect is not statistically significant for \CIV.  Spearman rank correlation coefficients ($\rho$) are given with the probability ($P$) of finding this distribution of points by chance listed underneath. \label{fig:fwhm}}
\end{figure*}

The scatter between black hole masses estimated from different emission lines that results from varying radio core dominance is derived largely from the broad-line FWHM dependence on radio core dominance.  Figure~\ref{fig:fwhm} shows log~$R$ versus FWHM for all three broad emission lines, where log~$R$ is on the y-axis for an easy comparison with previous results.  We confirm that \Hb\ has the strongest log~$R$ dependence, followed by \MgII, and finally \CIV\ which shows no dependence on radio core dominance.  The given correlation coefficients are Spearman rank correlation coefficients and the probability of finding these distributions of points by chance are listed in each panel.

\subsection{The orientation dependence of black hole masses}
The radio core dominance dependence of broad-line FWHM for \Hb\ and \MgII\ implies that black hole masses estimated from those lines may also have a similar dependence \citep{baskin05,jarvis06}.  \citet{tang12} show that there is scatter between black hole masses estimated from different broad emission lines.  We quantify the scatter around the one-to-one relationship by taking the standard deviation of the residuals in the y direction from the best fit to the parameters: around \Hb\ and \CIV\ is 0.40 dex, around \MgII\ and \CIV\ is 0.194 dex, and around \Hb\ and \MgII\ is 0.20 dex.  This scatter is in part due to radio core dominance and can be reduced by applying a correction based on log~$R$.
 
\subsubsection{Correlations}
In order to investigate how different black hole mass scaling relationships might depend on radio core dominance, we correlate the difference between the log of black hole masses estimated from different scaling relationships with radio core dominance.  Table~\ref{tab:correlation} gives the Spearman rank correlation coefficients and corresponding two-tailed probabilities.  We normalize the masses rather than simply correlating masses derived from \Hb, \MgII, and \CIV\ individually with radio core dominance because we want to isolate the bias in black hole mass from orientation and avoid effects associated with the sample having a range in intrinsic black hole mass, which may mask orientation effects.  The caveat to this approach is that if any systematic differences between the line width measurements depend on orientation, they will be folded into our results.

\input{correlations.tex}

Table~\ref{tab:correlation} indicates that the log~$R$ dependence of \Hb\ and \MgII\ FWHM is propagated to the black hole masses derived from the \Hb\ and \MgII\ scaling relationships.  Taking into account the number of objects available for each correlation and the probability indicated by the Spearman rank correlation test, the difference between the log of black hole masses estimated from \Hb\ and \CIV, \MgII\ and \CIV, and \Hb\ and \MgII\, is correlated with radio core dominance with a t-ratio of 3.5, 2.4, and 2.7, respectively.  This correlation with radio core dominance causes a range in the difference between black hole masses estimated from different emission lines that covers nearly an order of magnitude from the most jet-on (large values of log~$R$) to the most edge-on (small values of log~$R$) objects.  In practice, the actual range will depend on the range of orientations present in any given sample and uncertainty in the orientation from log~$R$.

Core and extended radio luminosities are also listed in Table~\ref{tab:correlation} in order to verify the source of the correlation between \Hb\ and \MgII-derived masses and radio core dominance.  The differences $\Delta$log$(M_{BH})($\Hb$-$\CIV$)$ and $\Delta$log$(M_{BH})($\MgII$-$\CIV$)$ are more significantly correlated with core luminosity than the extended luminosity, indicating that the correlations between the mass difference and log~$R$ are due to variation in the core luminosity among objects that are intrinsically similar.

\subsubsection{Regression analysis}
We also performed a regression analysis in order to describe the trends with radio core dominance.  The fits in this paper are made by minimizing the chi-squared statistic using \textsc{mpfit} \citep{markwardt09} which employs the Levenberg-Marquardt least squares method.  The uncertainties in log~$R$ are considered to be negligible compared to those in black hole mass because the fluxes are well known (although the uncertainty in $\theta$ is large) unless otherwise noted.

\begin{figure}
\begin{center}
\includegraphics[width=8.9 truecm]{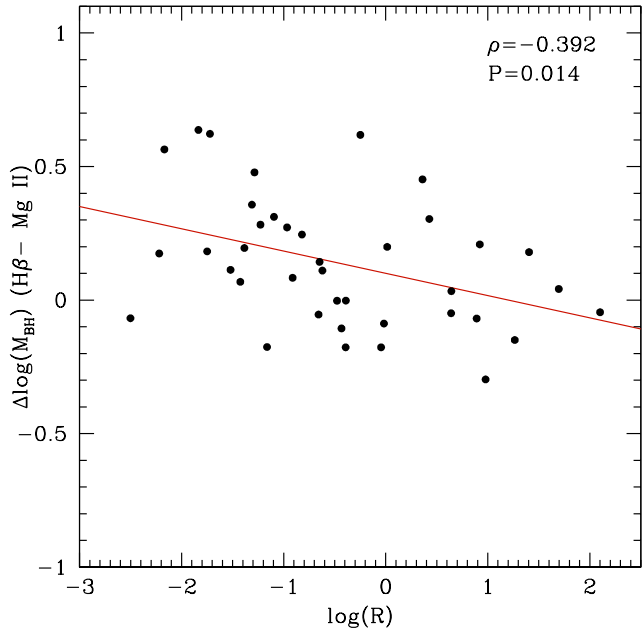}
\end{center}
\caption{Log \mbhhb $-$log \mbhmgii\ versus log~$R$ for 39 objects where measurements are available for all three quantities.  The solid red line shows the linear regression with a nonzero intercept.  The dependence on log~$R$ is weaker; the relationship is significant at nearly the 3 $\sigma$ level.}
\label{fig:HbMgIIfit}
\end{figure}

\begin{figure}
\begin{center}
\includegraphics[width=8.9 truecm]{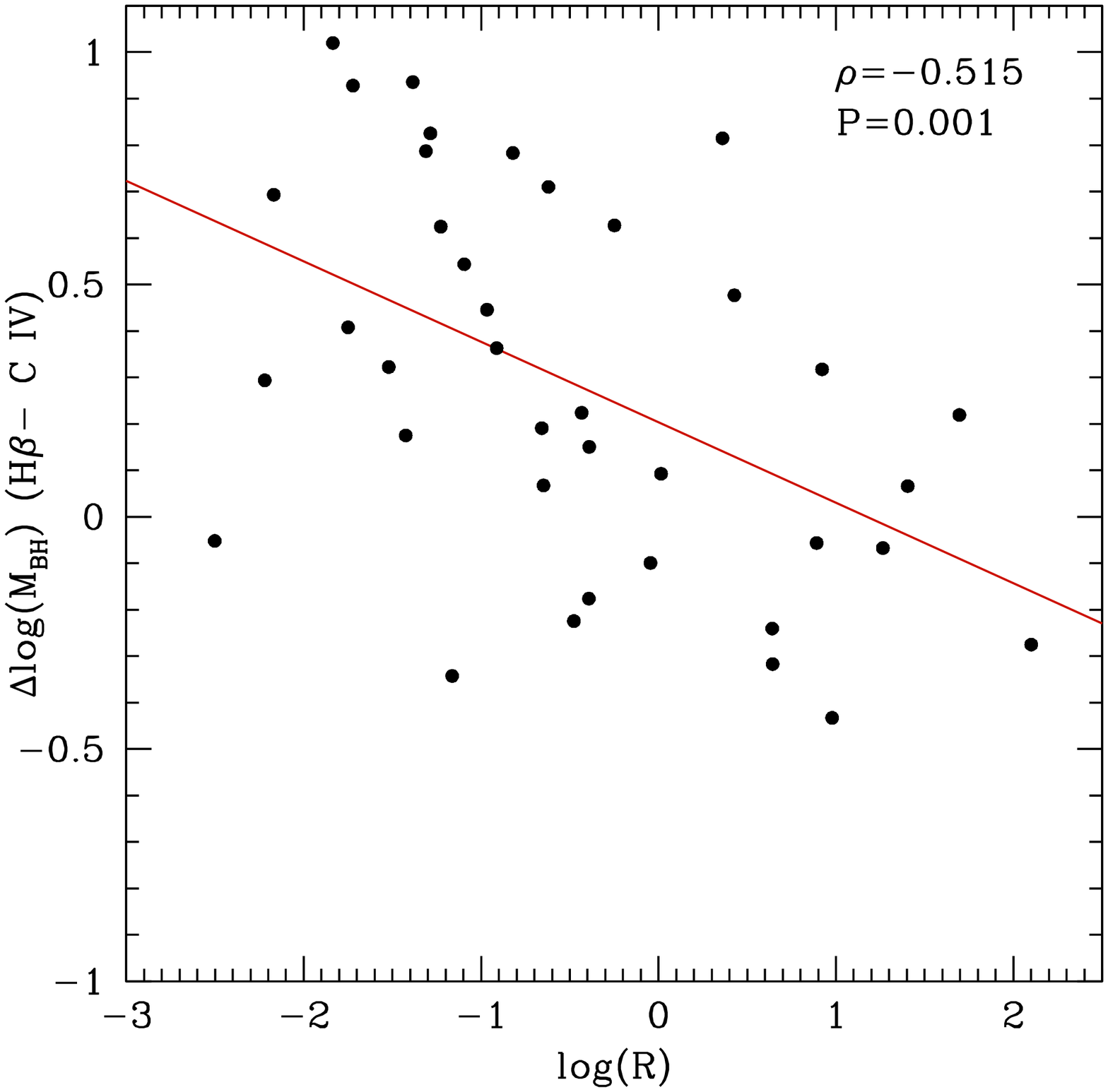}
\end{center}
\caption{Log \mbhhb $-$log \mbhciv\ versus log~$R$ for 38 objects where measurements are available for all three quantities.  The solid red line shows the linear regression with a nonzero intercept.  The relationship is significant at greater than the 3 $\sigma$ level.}
\label{fig:HbCIVfit}
\end{figure}

\begin{figure}
\begin{center}
\includegraphics[width=8.9 truecm]{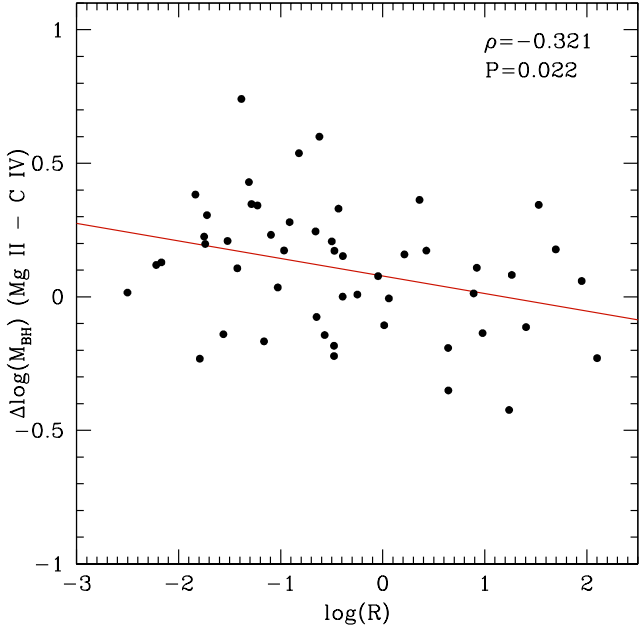}
\end{center}
\caption{Log \mbhmgii $-$log \mbhciv\ versus log~$R$ for 51 objects where measurements are available for all three quantities.  The solid red line shows the linear regression with a nonzero intercept.  The relationship is significant at greater than the 2 $\sigma$ level.}
\label{fig:CIVMgIIfit}
\end{figure}

The results of the regression analysis also suggest that the \Hb\ black hole mass scaling relationship has the strongest dependance on radio core dominance, though \MgII\ also has a dependence, albeit weaker.  Figures~\ref{fig:HbMgIIfit}, \ref{fig:HbCIVfit}, and \ref{fig:CIVMgIIfit} show three combinations of difference in log of black hole mass versus log~$R$ with the regression line over-plotted.  The difference $\Delta$log$(M_{BH})($\Hb$-$\CIV$)$ shows the steepest slope versus log~$R$, indicating that \Hb\ has the strongest bias with radio core dominance.  The difference $\Delta$log$(M_{BH})($\MgII$-$\CIV$)$ has a slightly shallower slope, indicating that \MgII\ has a weaker dependence on log~$R$ than does \Hb.  The \Hb\ and \MgII-derived masses both depend on radio core dominance, but the negative slope of $\Delta$log$(M_{BH})($\Hb$-$\MgII$)$ versus log~$R$ supports the conclusion that \Hb\ has the stronger dependence.

The significance of this effect can be increased by considering the one-tailed probabilities for the correlation coefficients.  Because we expect \Hb\ scaling relationships to underestimate black hole masses compared to \CIV\ and \MgII\ scaling relationships for jet-on objects, we can halve the probabilities given in Table~\ref{tab:correlation}.  According to the one-tailed probabilities, the difference between the log of black hole mass versus log~$R$ is correlated with a probability of finding the observed distribution of points by chance of: $<0.7\%$ for \Hb\ and \MgII\, $<0.05\%$ for \Hb\ and \CIV\, and $<1.1\%$ for \MgII\ and \CIV.

%The significance may be lower than is indicated by a Spearman rank correlation test because black hole masses estimated from different emission lines are correlated.  A Monte-Carlo simulation which takes into account the co-dependence of the black hole masses yields lower significance, the difference in masses estimated from \Hb\ and \CIV\ becomes significant at the $2\sigma$ level and \MgII\ and \CIV\ at $1\sigma$.  Regardless, the effect is derived from the dependence of FWHM on radio core dominance which is very clear in Figure~\ref{fig:fwhm}. 

\subsubsection{Luminosity considerations}
\begin{figure*}
\begin{minipage}[!b]{5cm}
\centering
\includegraphics[width=6cm]{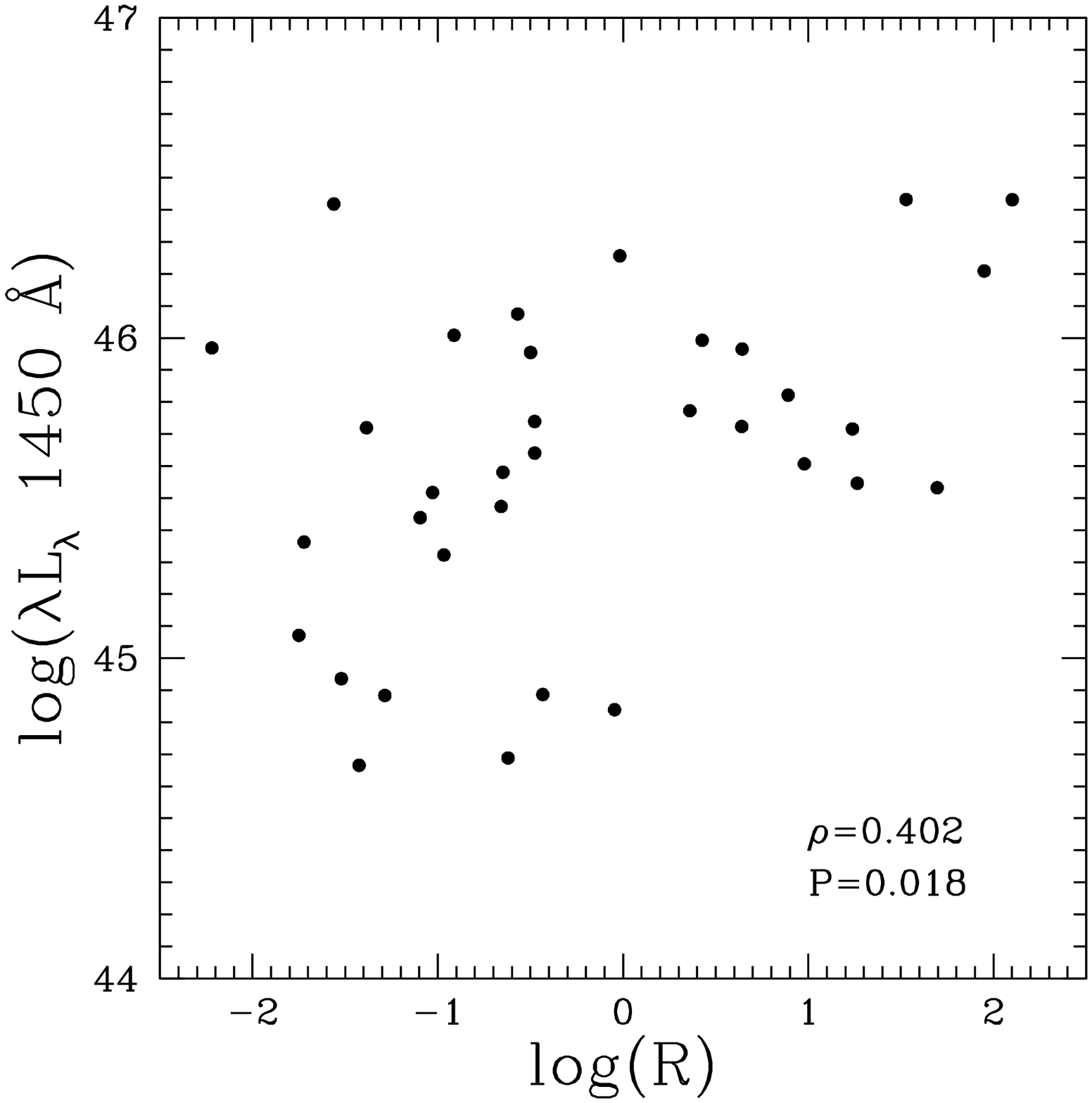}
\end{minipage}
\hspace{0.6cm}
\begin{minipage}[!b]{5cm}
\centering
\includegraphics[width=6cm]{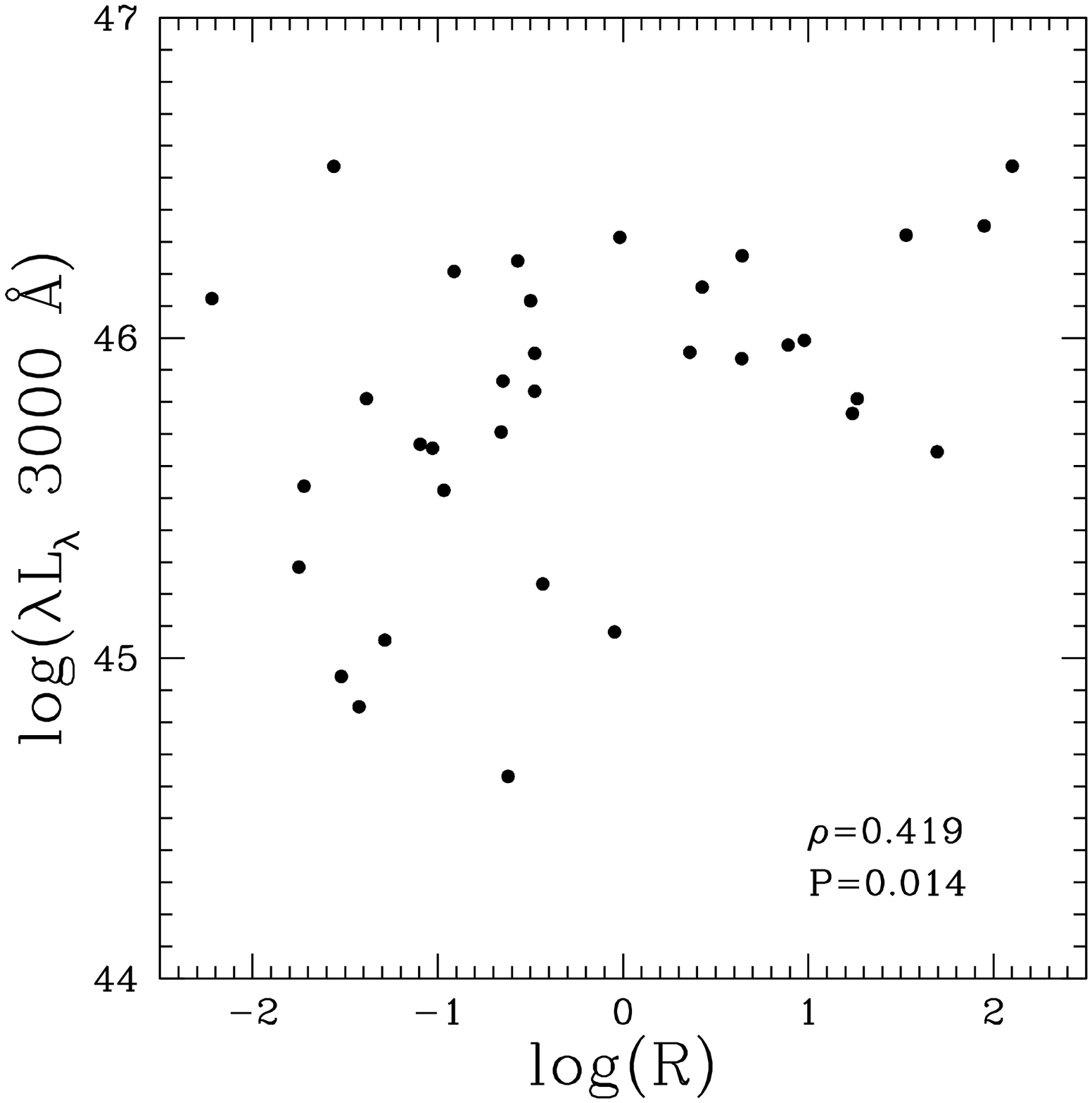}
\end{minipage}
\hspace{0.6cm}
\begin{minipage}[!b]{5cm}
\centering
\includegraphics[width=6cm]{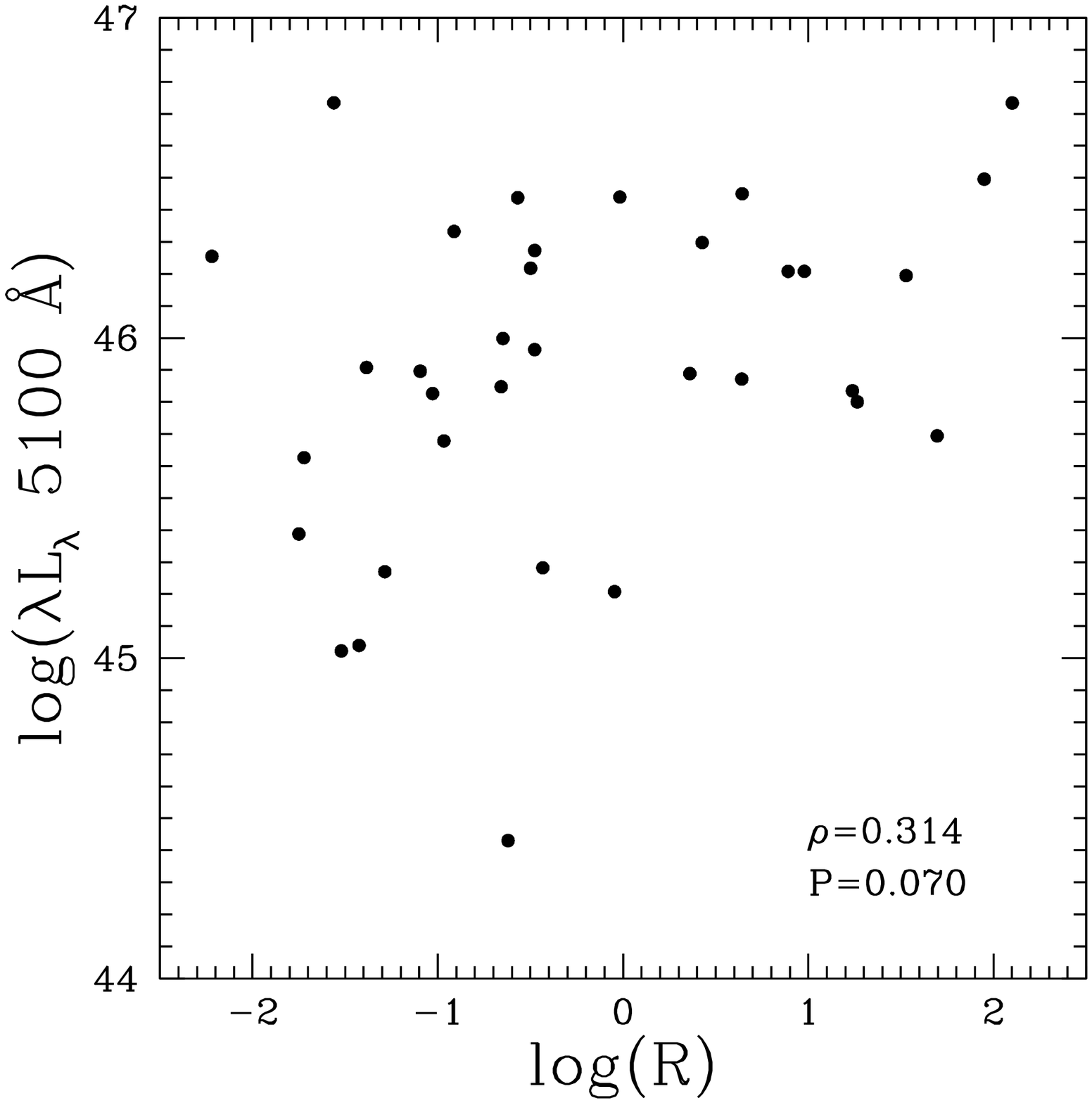}
\end{minipage}                             
\caption{$\lambda$L$_{\lambda}$ at 1450, 3000, and 5100 \AA\ versus log~$R$.  The correlations between the difference in the log of black hole mass and log~$R$ occur in spite of the correlations between $\lambda$L$_{\lambda}$ and log~$R$.  The Spearman rank correlation coefficient and corresponding probability of finding this distribution of points by chance are given.}
\label{fig:lr}
\end{figure*}

In order to verify that the dependence of difference in the log of black hole masses from different scaling relationships on radio core dominance is not the result of a luminosity effect which enters via the $R-L$ relationship, we consider our sample in the luminosity-log~$R$ plane.  We expect to see a slight correlation because, although the selection for similar extended luminosities means that we are looking at similar luminosity objects in principle, orientation effects are likely to introduce a trend into the observed luminosity \citep[e.g,][]{nb10} because accretion discs are not isotropic emitters.  Figure~\ref{fig:lr} shows an increase in $\lambda$L$_{\lambda}$ with radio core dominance for our objects.  More face-on objects (larger log~$R$ values) have higher apparent luminosities, which will result in higher calculated black hole masses for these objects because luminosity goes to the one half power in the mass scaling relationships.  We observe the \citet{nb10} effect and the dependence of difference in black hole mass on radio core dominance exists in spite of it.

In principle, continuum luminosity and FWHM must both be corrected for inclination and an empirical correction to black hole mass will incorporate both dependancies, though the primary effect is derived from the FWHM dependence, which enters as the square in black hole mass scaling relationships while luminosity is only to the one half power.  

In practice, we find that the empirical mass correction does not account for the luminosity dependence because we fit the correction based on the differences in the log of the black hole mass, so that the luminosity also enters as the difference, $\Delta L(5100 \textrm{\AA}-1450\textrm{\AA})$ and $\Delta L(3000\textrm{\AA}-1450\textrm{\AA})$. Figure~\ref{fig:lr} shows that the optical and UV luminosities have nearly the same dependence on log~$R$, so these differences are flat with radio core dominance.  In order to remove the radio core dominance dependence from the luminosity for individual single-epoch mass estimates, we therefore provide corrections in Equations~\ref{eqn:L1450}-\ref{eqn:L5100}.\\

\noindent At 1450 \AA:
\input{L1450.tex}
\noindent At 3000 \AA:
\input{L3000.tex}
\noindent At 5100 \AA:
\input{L5100.tex}

$\lambda$L$_{\lambda}$ is the uncorrected monochromatic luminosity and $\lambda$L$^{\prime}_{\lambda}$ is the luminosity corrected for log~$R$.  The broken fit around log~$R=0$ is naturally motivated by the shape of the data in Figure~\ref{fig:lr}.

The ratio of radio core luminosity at 5 GHz to a monochromatic optical luminosity, log $R_{V}$ (different from radio loundess parameter, log $R^{*}$) has been suggested to be a better indication of orientation than the radio core dominance, log~$R$, with some caveats \citep{willsbro95}.  As with \citet{vestergaard00}, our result persists, but is slightly less significant, when performing the analysis with log $R_{V}$ instead of log~$R$.

\subsection{Correcting black hole masses for orientation}
The correlations presented demonstrate a likely orientation dependence that is propagated into black hole masses calculated from \Hb\ and \MgII\ FWHM measurements.  The fact that no significant radio core dominance dependence is observed for the FWHM of \CIV\ allows a correction for orientation to \Hb\ and \MgII-derived masses.  For \Hb, we used the fit of $\Delta$log$(M_{BH})($\Hb$-$\CIV$)$ versus log~$R$ to zero the difference between \Hb\ and \CIV-derived black hole masses, normalizing at log~$R=0$.  We note that other normalizations are possible, and an edge-on normalization might be preferable but is not possible here as a true edge-on view is likely obscured by a dusty torus.  With the orientation dependence removed from the difference, it can be combined with the orientation unbiased \CIV\ black hole mass estimates to give an unbiased estimate of \Hb\ black hole mass.  This process was repeated for \MgII.  Equations~\ref{eqn:Hbnew} and \ref{eqn:MgIInew} can be used to estimate black hole masses unbiased by orientation from \Hb\ or \MgII.

\input{mbhfix.tex}

$M_{BH}(\textrm{\Hb})^{\prime}$ is the log of the \Hb-based black hole mass with the log~$R$ dependence removed, and $M_{BH}(\textrm{\Hb})$ is the log of the orientation-biased, \Hb-based black hole mass estimated from the scaling relationship.  The same conventions apply for \MgII.  The uncertainties given here are the 1-$\sigma$ errors in the fit coefficients and are not independent.  

\begin{figure*}
\begin{minipage}[!b]{8cm}
\centering
\includegraphics[width=8cm]{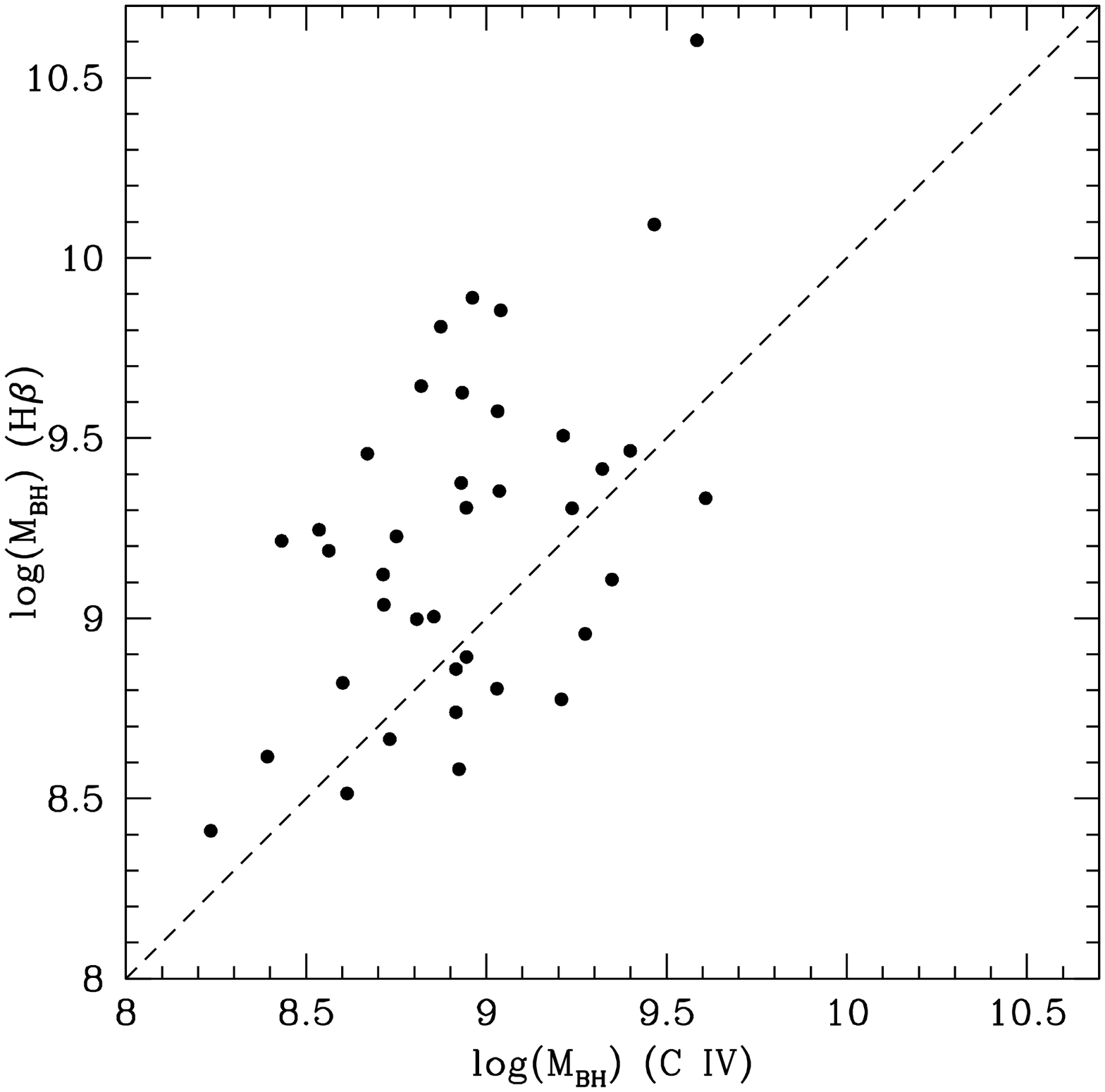}
\end{minipage}\hspace{0.6cm}
\hspace{0.6cm}
\begin{minipage}[!b]{8cm}
\centering
\includegraphics[width=8cm]{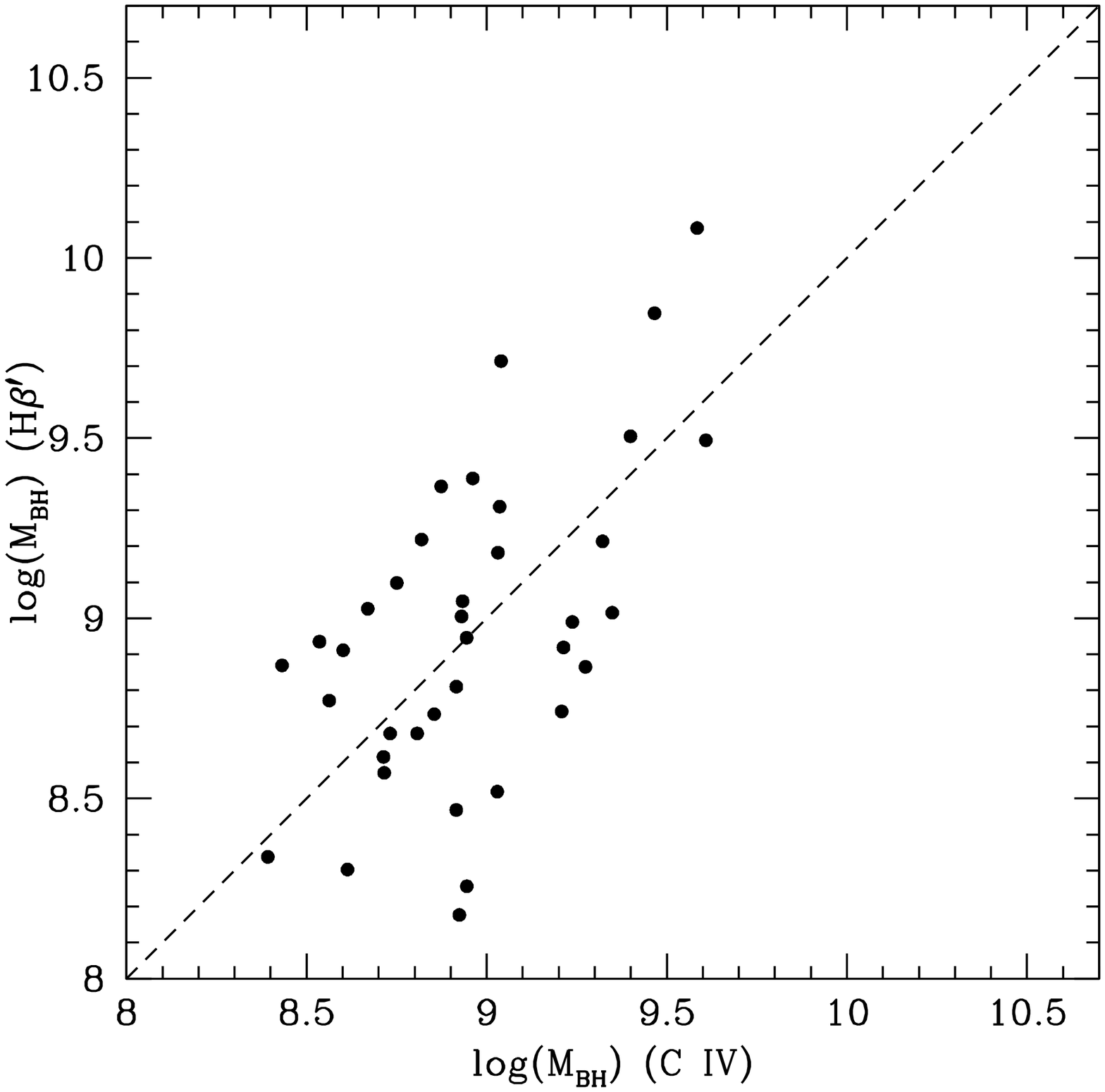}
\end{minipage}
\hspace{0.6cm}                   
\caption{\Hb\ versus \CIV-based black hole masses (left) and corrected \Hb$^{\prime}$ versus \CIV-based black hole masses (right).  The dashed line is the one-to-one relationship. \label{fig:Hbscatter}}
\end{figure*}

\begin{figure*}
\begin{minipage}[!b]{8cm}
\centering
\includegraphics[width=8cm]{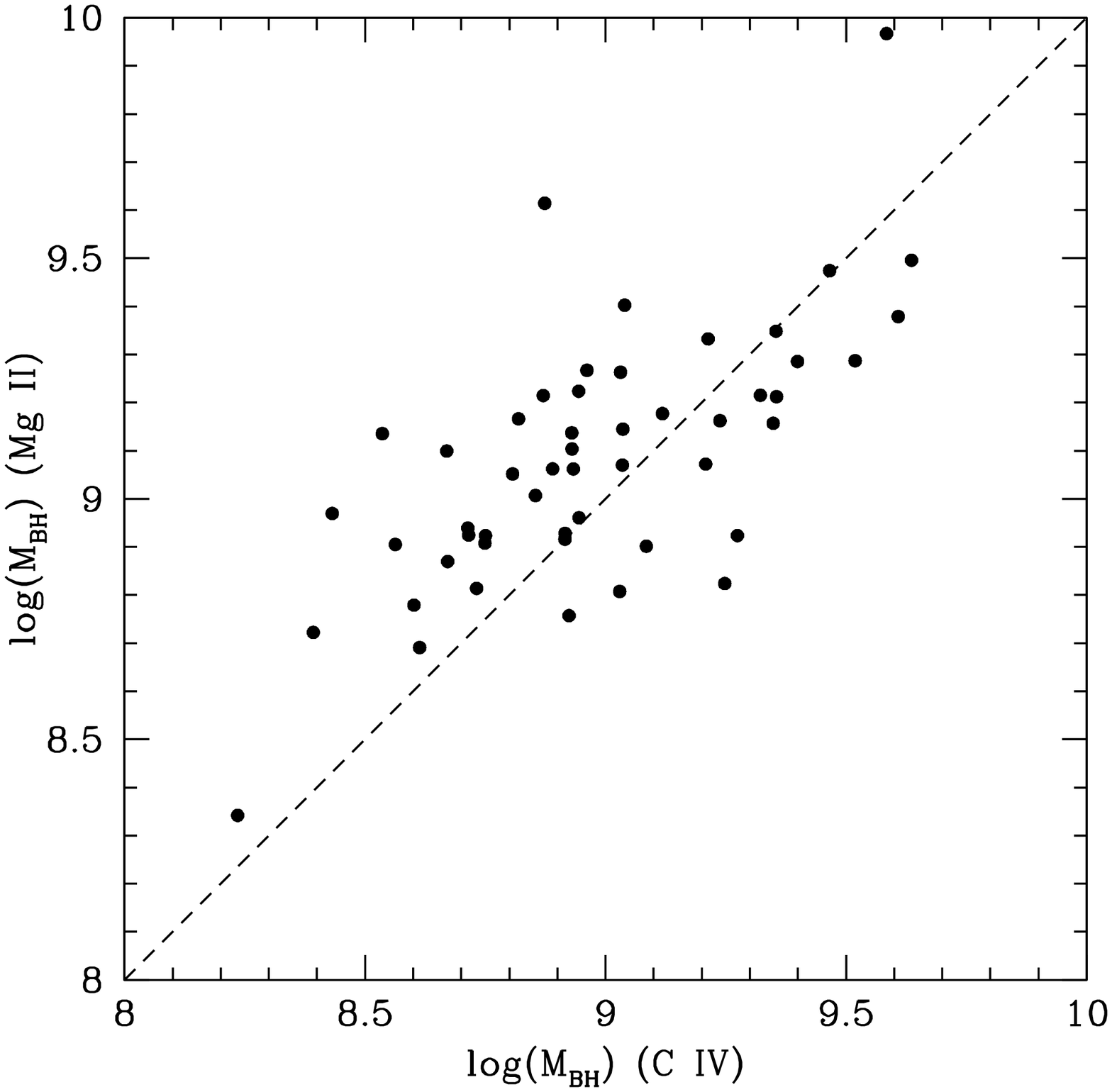}
\end{minipage}\hspace{0.6cm}
\hspace{0.6cm}
\begin{minipage}[!b]{8cm}
\centering
\includegraphics[width=8cm]{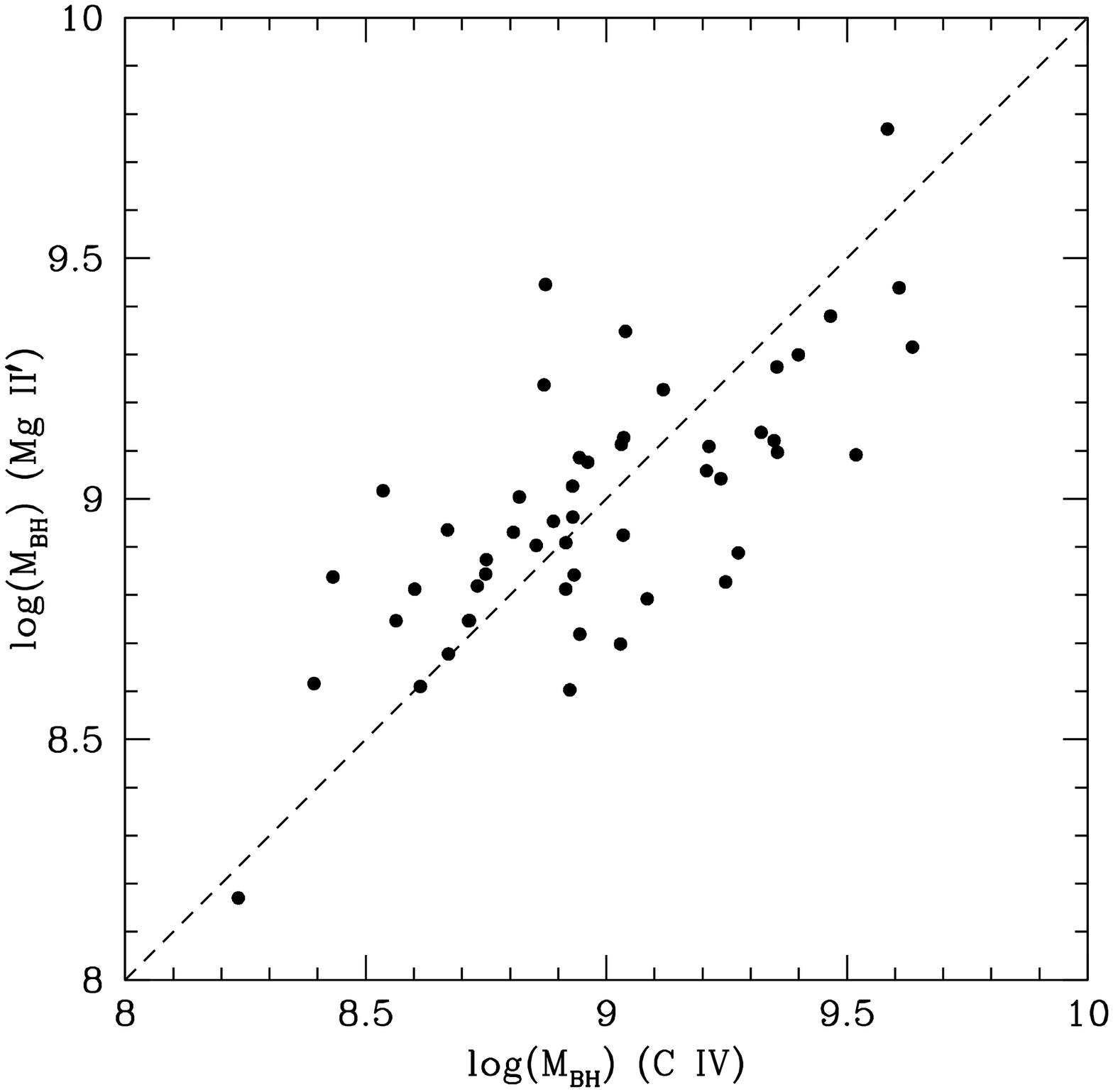}
\end{minipage}
\hspace{0.6cm}                   
\caption{\MgII\ versus \CIV-based black hole masses (left) and corrected \MgII$^{\prime}$ versus \CIV-based black hole masses (right).  The dashed line is the one-to-one relationship. Note that, while the reduction in scatter is modest, the points now fall more evenly around the one-to-one line.  \label{fig:MgIIscatter}}
\end{figure*}

Applying this correction reduces the scatter between black hole masses estimated from different broad emission lines.  The scatter around the one-to-one relationship is reduced from $0.40$ dex to $0.35$ dex for \Hb\ and \CIV, $0.194$ dex to $0.188$ dex for \MgII\ and \CIV, and $0.20$ to $0.18$ for \Hb\ and \MgII.  Figures~\ref{fig:Hbscatter} and \ref{fig:MgIIscatter} show the reduction in scatter that results from applying the corrections for log~$R$ for \Hb\ and \MgII.  Original masses are shown in the left panel and corrected masses are on the right.

The caveat for these corrections is that they are linear fits to data that display a linear relationship in log space with large scatter.  The errors are large due to the large scatter present in the data, and fitting with a different algorithm or prescription will yield a slightly different correction. 

% Radio spectral index
%%%%%%%%%%%%%%%%%%%%%%%%%%%%%%%%%%%%%%%%%%%%%%%%%%%%%%%%%%%%%%%%%%%%%%%%%%%%%%%%%

\subsection{Radio spectral index}
We repeat the analysis for radio spectral index, an orientation indicator that is more easily measured than radio core dominance because the object does not need to be resolved in the radio image.  We present only the results for radio spectral index measured from all the available radio data.  The results for radio spectral index measured between 1.4 and 4.9 GHz are similar, but the correlations are slightly weaker.

\begin{figure}
\begin{center}
\includegraphics[width=8.9 truecm]{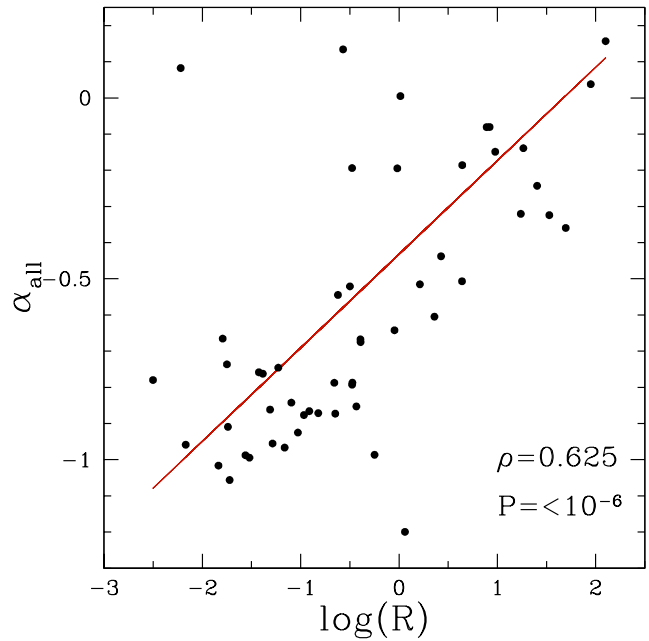}
\end{center}
\caption{Radio spectral index measured for all available radio data points versus radio core dominance.  $\alpha_{all}$ is a good proxy for log~$R$, which is traditionally used as an orientation indicator.  The biggest outlier in this correlation has a spectral shape in the radio that is not well fit by a power law.}
\label{fig:orient}
\end{figure}

Radio spectral index is a good proxy for radio core dominance, the two are correlated with a probability of less than $10^{-6}$ for the correlation to arise by chance.  Figure~\ref{fig:orient} shows radio spectral index measured from all the available radio data points versus radio core dominance.  We find a best-fitting line of:

\input{orient.tex}

\begin{figure*}
\begin{minipage}[!b]{5cm}
\centering
\includegraphics[width=6cm]{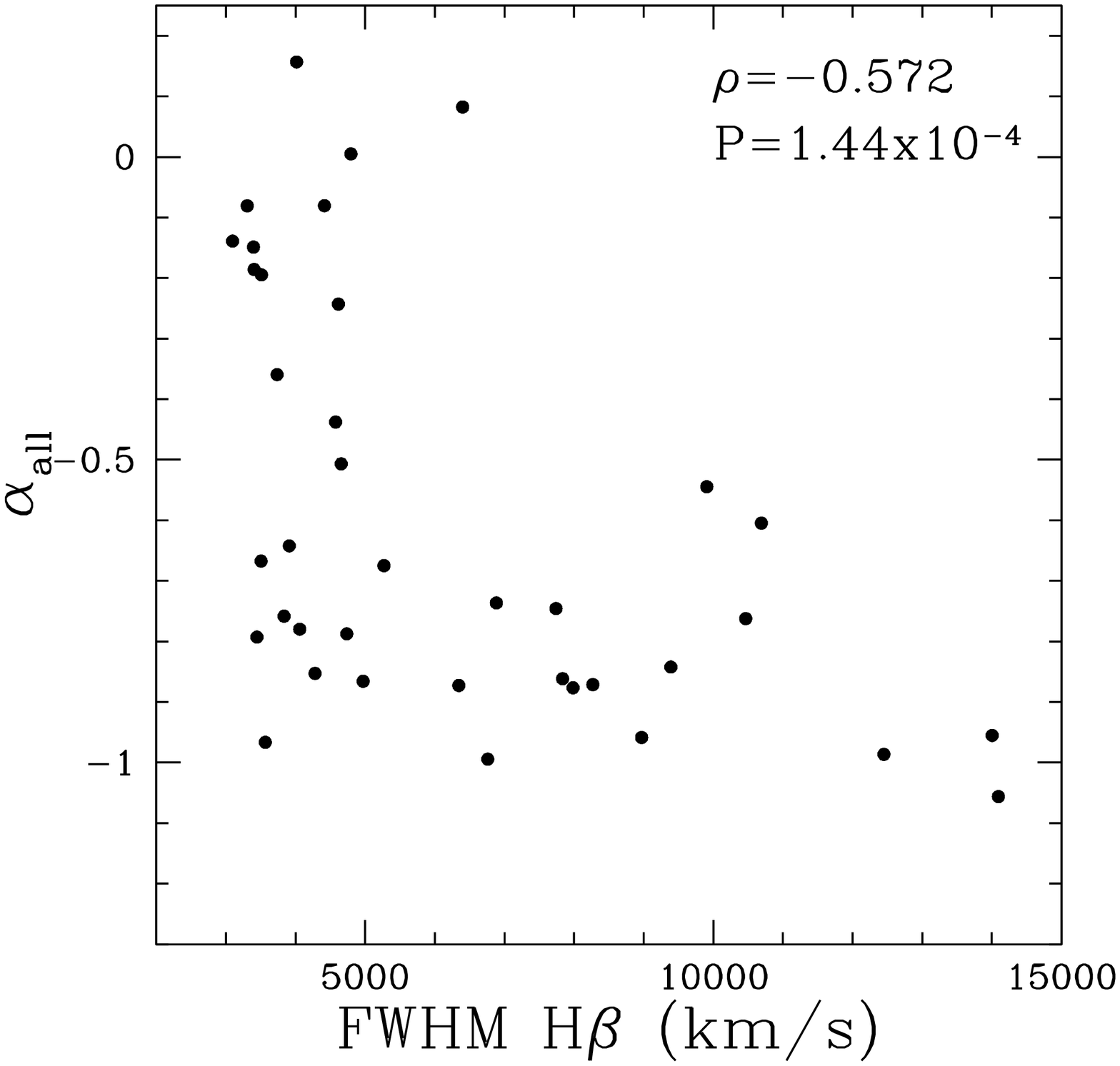}
\end{minipage}
\hspace{0.6cm} 
\begin{minipage}[!b]{5cm}
\centering
\includegraphics[width=6cm]{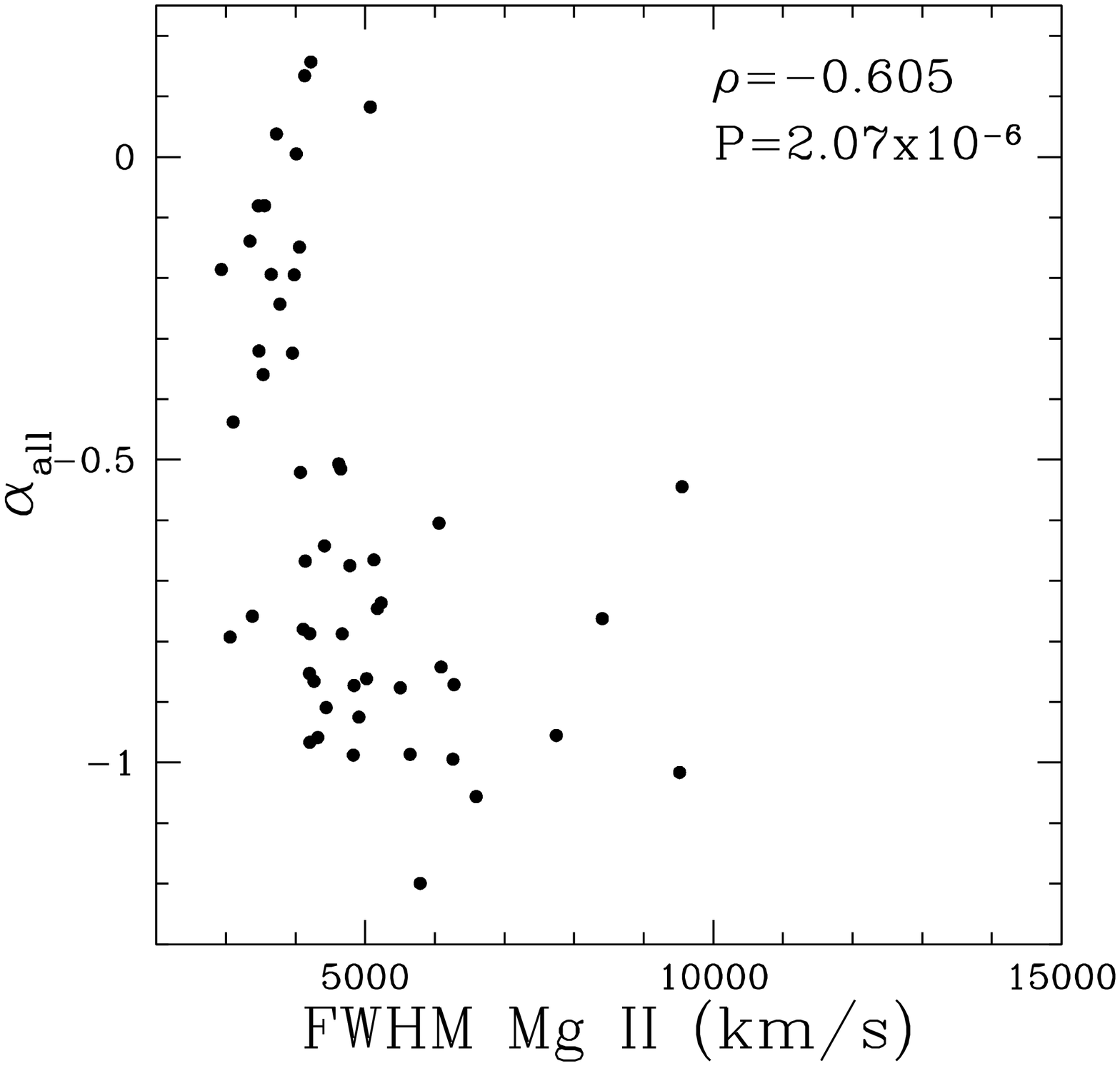}
\end{minipage}
\hspace{0.6cm}
\begin{minipage}[!b]{5cm}
\centering
\includegraphics[width=6cm]{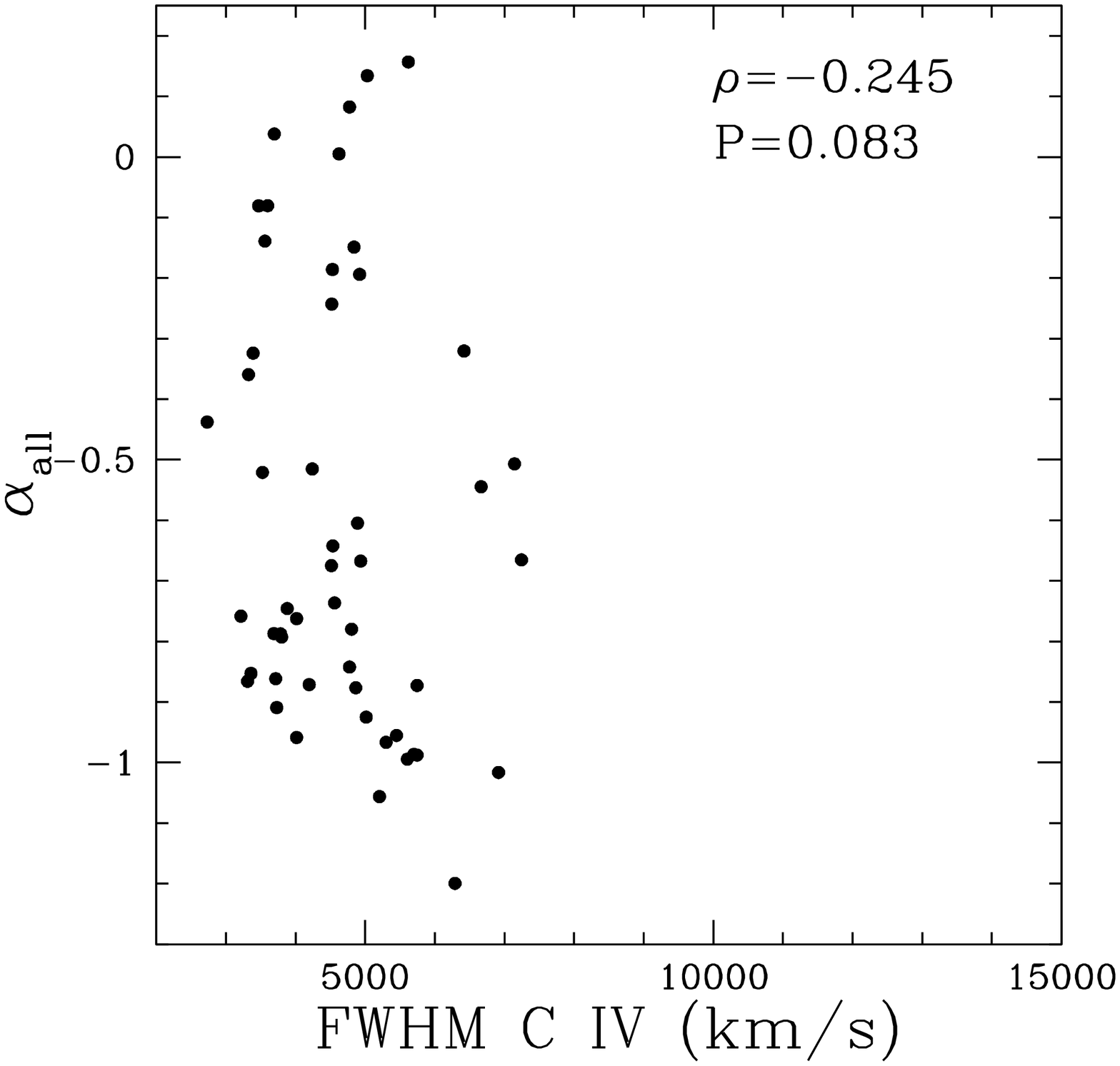}
\end{minipage}                             
\caption{Radio spectral index versus FWHM for \Hb, \MgII, and \CIV.  The effect seen with radio core dominance is reproduced for radio spectral index.  Spearman rank correlation coefficients are given with the probability of finding this distribution of points by chance listed underneath. \label{fig:fwhm_alpha}}
\end{figure*}

The orientation dependence of FWHM of \Hb\ and \MgII\ and lack of a dependence for \CIV\ that is indicated by a radio core dominance dependence is also present for radio spectral index.  Figure~\ref{fig:fwhm_alpha} shows this dependence on radio spectral index for all three lines.

In order to investigate how different black hole mass scaling relationships might depend on radio spectral index, we correlate it with the difference between the log of black hole masses estimated from different scaling relationships.  The Spearman rank correlation coefficients and corresponding two-tailed probabilities are given in Table~\ref{tab:correlation}.

Taking into account the number of objects available for each correlation and the probability indicated by the Spearman rank correlation test, the difference between the log of black hole masses estimated from \Hb\ and \CIV, \MgII\ and \CIV, and \Hb\ and \MgII\, is correlated with radio spectral index with a t-ratio of 3.3, 2.5, and 2.85, respectively.  As with radio core dominance, the correlation with radio spectral index causes a range in the difference between black hole masses estimated from different emission lines that covers about an order of magnitude from the most jet-on (large values of $\alpha_{all}$) to the most edge-on (small values of $\alpha_{all}$) objects.  Similarly, in practice, the actual range will depend on the range of orientations present in any given sample and uncertainty in the orientation from $\alpha_{all}$.

\begin{figure}
\begin{center}
\includegraphics[width=8.9 truecm]{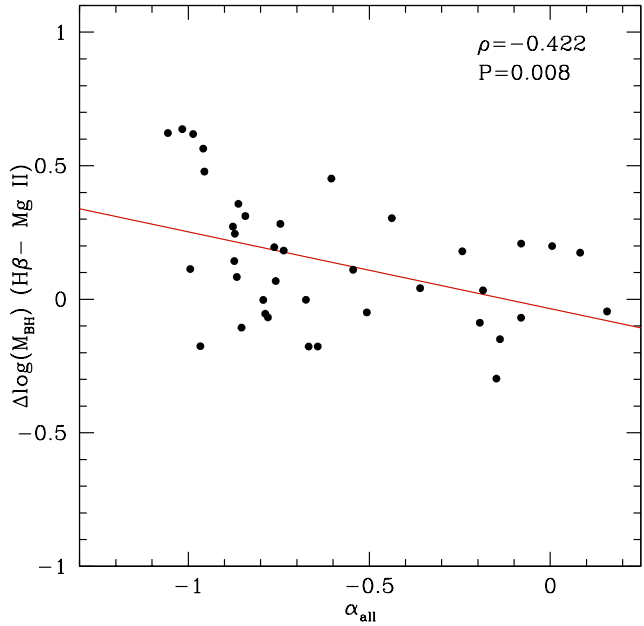}
\end{center}
\caption{Log \mbhhb $-$log \mbhmgii\ versus radio spectral index for 39 objects where measurements are available for all three quantities.  The solid red line shows the linear regression with a nonzero intercept.  The relationship is significant at approximately the 3 $\sigma$ level.}
\label{fig:HbMgIIfit_alpha}
\end{figure}

\begin{figure}
\begin{center}
\includegraphics[width=8.9 truecm]{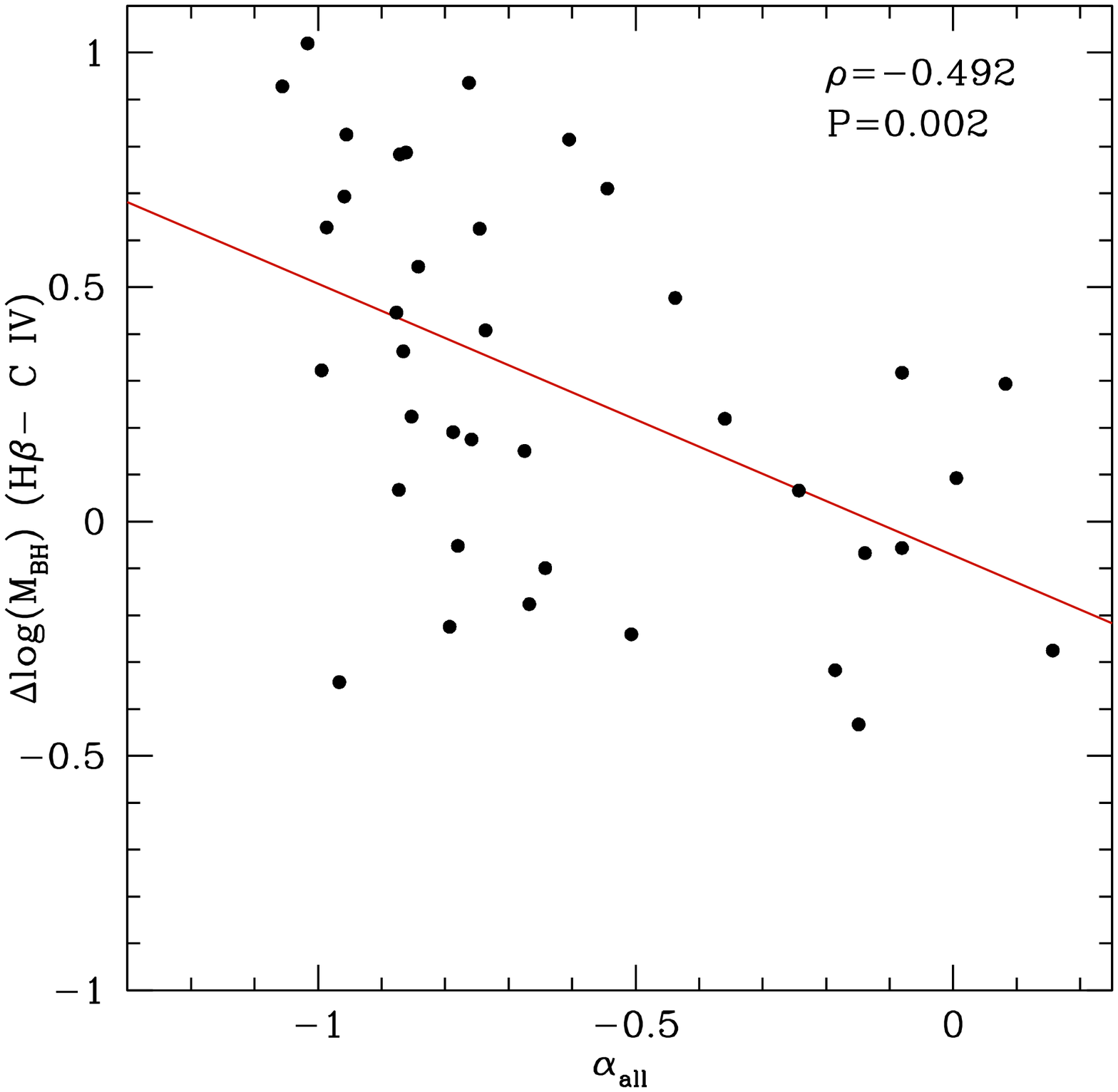}
\end{center}
\caption{Log \mbhhb $-$log \mbhciv\ versus radio spectral index for 38 objects where measurements are available for all three quantities.  The solid red line shows the linear regression with a nonzero intercept.  The relationship is significant at greater than the 3 $\sigma$ level.}
\label{fig:HbCIVfit_alpha}
\end{figure}

\begin{figure}
\begin{center}
\includegraphics[width=8.9 truecm]{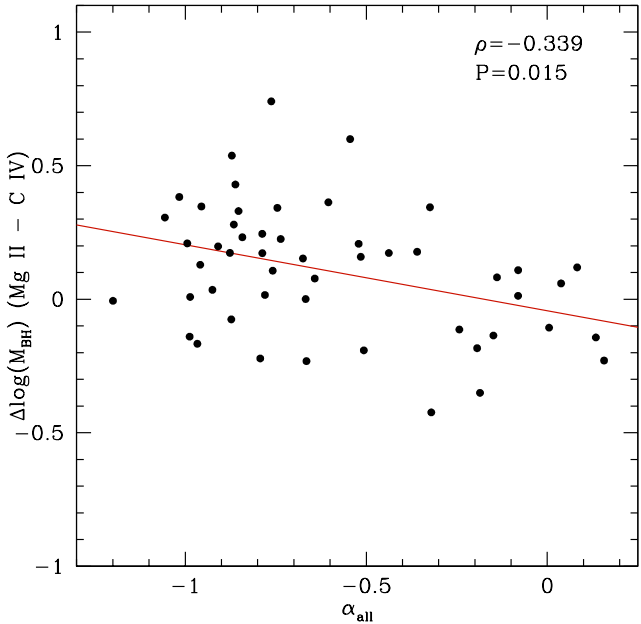}
\end{center}
\caption{Log \mbhmgii $-$log \mbhciv\ versus radio spectral index for 51 objects where measurements are available for all three quantities.  The solid red line shows the linear regression with a nonzero intercept.  The relationship is significant at nearly the 3 $\sigma$ level.}
\label{fig:CIVMgIIfit_alpha}
\end{figure}

We also repeated the regression analysis in order to describe the trends with radio spectral index.  The uncertainties in $\alpha_{all}$ are considered to be negligible compared to those in black hole mass.  The trends are very similar to those for radio core dominance.  Figures~\ref{fig:HbMgIIfit_alpha}, \ref{fig:HbCIVfit_alpha}, and \ref{fig:CIVMgIIfit_alpha} show three combinations of difference in log of black hole mass versus $\alpha_{all}$ with the regression line over-plotted.

\begin{figure*}
\begin{minipage}[!b]{5cm}
\centering
\includegraphics[width=6cm]{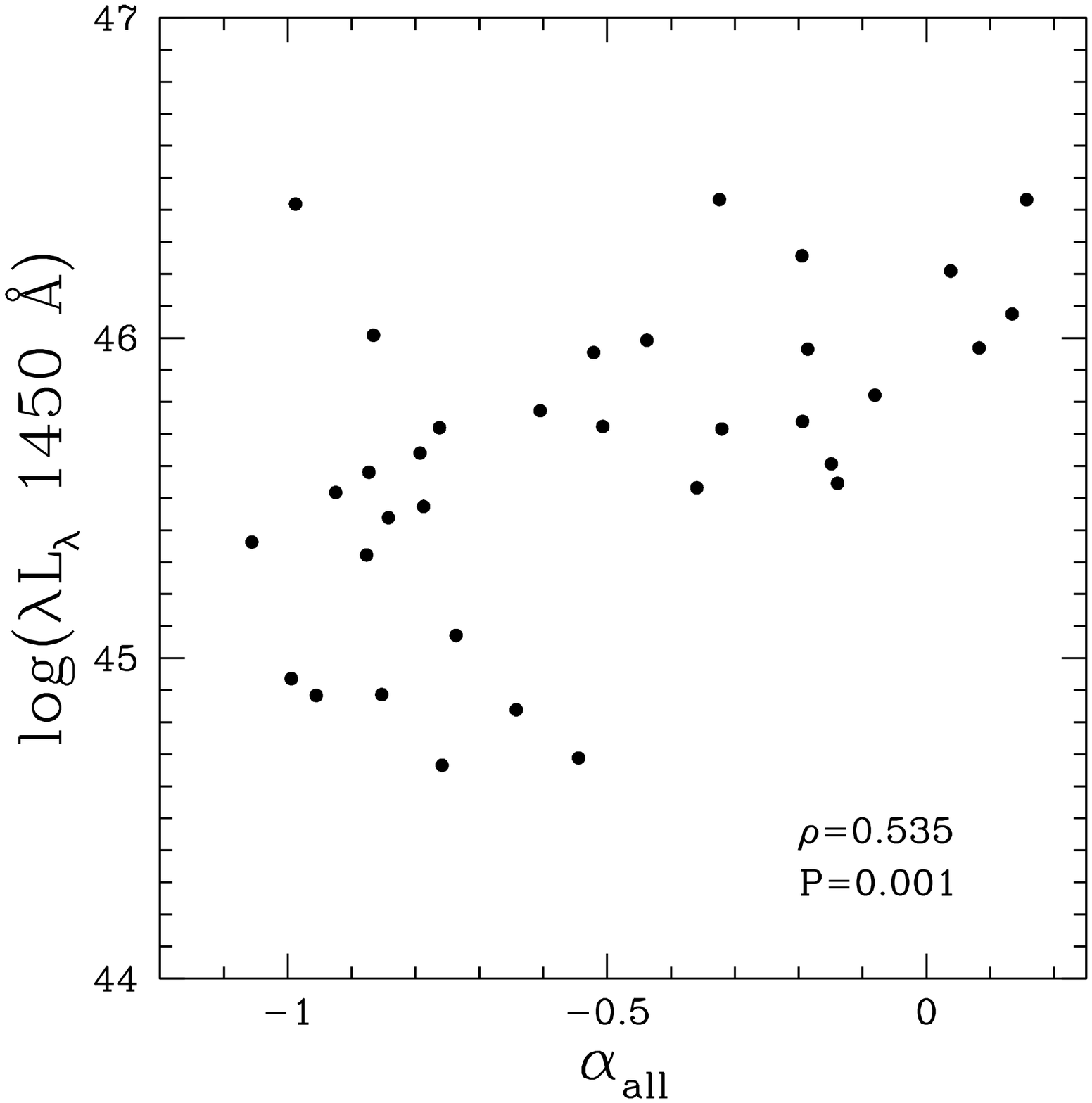}
\end{minipage}
\hspace{0.6cm}
\begin{minipage}[!b]{5cm}
\centering
\includegraphics[width=6cm]{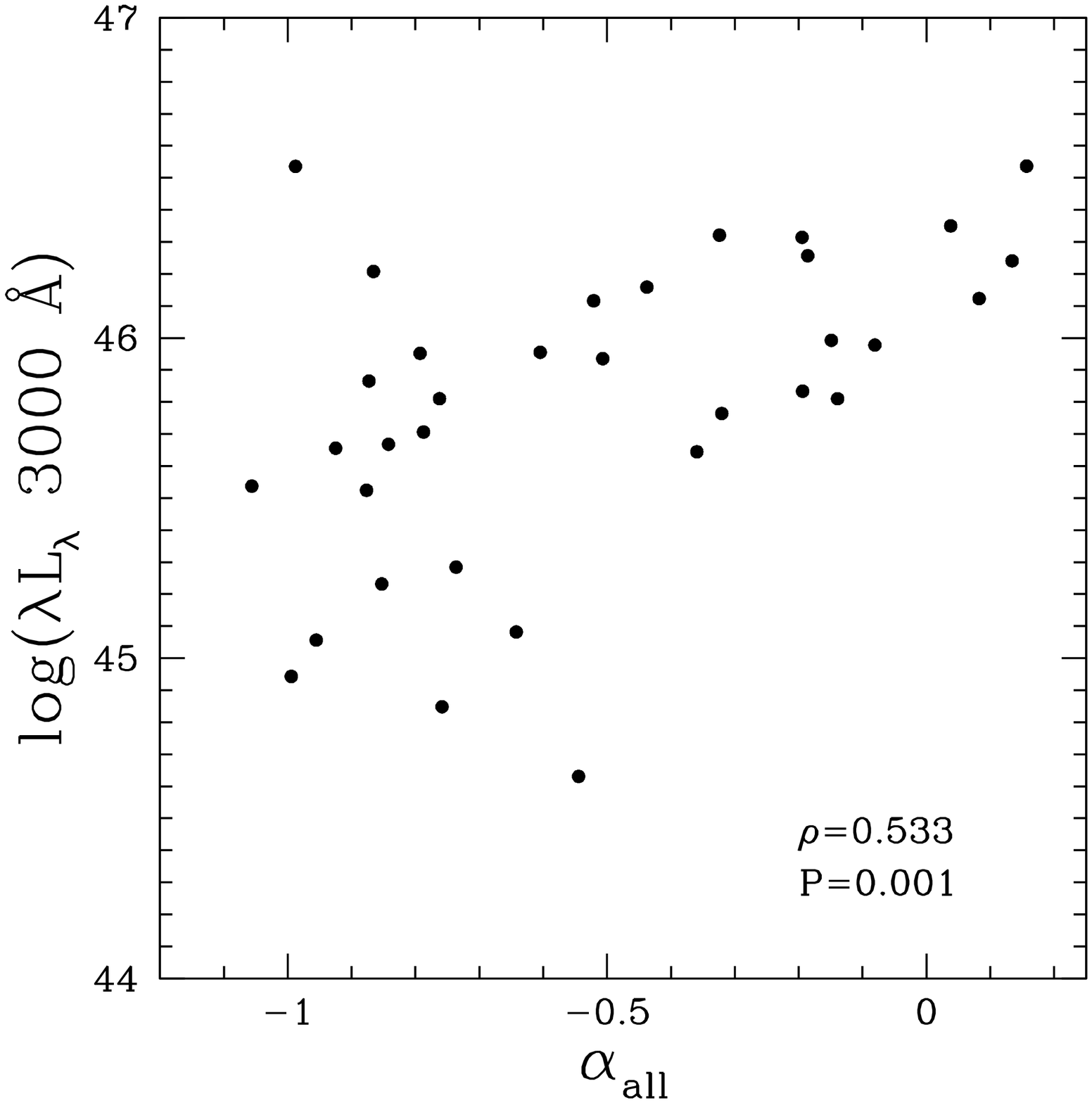}
\end{minipage}
\hspace{0.6cm}
\begin{minipage}[!b]{5cm}
\centering
\includegraphics[width=6cm]{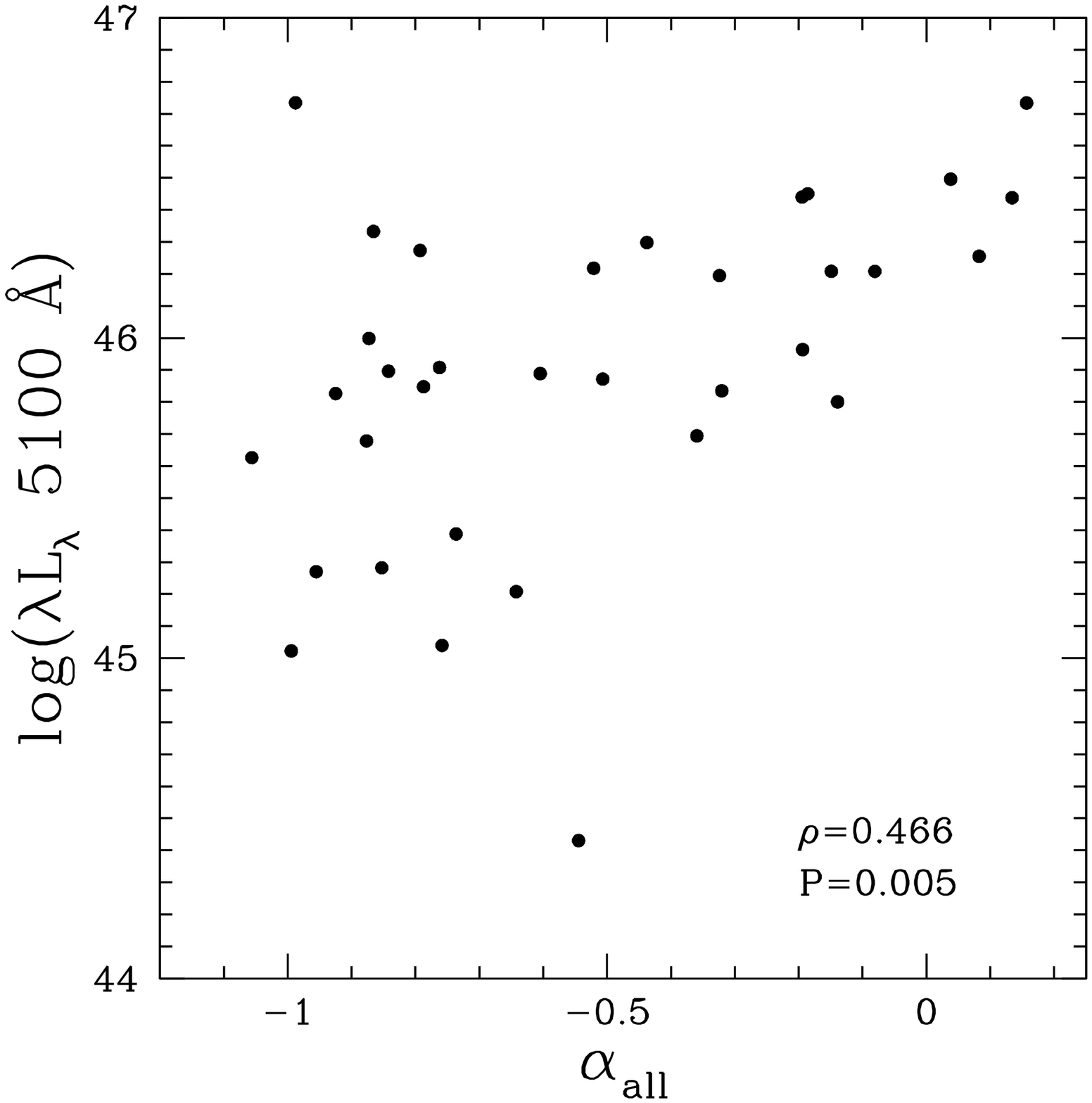}
\end{minipage}                             
\caption{$\lambda$L$_{\lambda}$ at 1450, 3000, and 5100 \AA\ versus $\alpha_{all}$.  The correlations between black hole mass difference and $\alpha_{all}$ occur in spite of the correlations between $\lambda$L$_{\lambda}$ and $\alpha_{all}$.  The Spearman rank correlation coefficient and corresponding probability of finding this distribution of points by chance are given.}
\label{fig:lr_alpha}
\end{figure*}

The behavior of the 1450, 3000, and 5100 \AA\ monochromatic luminosities with radio spectral index is similar as for radio core dominance.  The differences in luminosity are not correlated with $\alpha_{all}$ but the individual luminosities are, so we provide a correction to these that should be applied before calculating black hole masses from the single-epoch scaling relationships.  There is no evidence in Figure~\ref{fig:lr_alpha} that a broken fit is necessary here, so we fit only one line.  The data have a lot of irregular scatter and we find that assuming the errors in radio spectral index are negligible and weighting all the monochromatic luminosities similarly gives the best fit.  We normalize the correction at $\alpha_{all}=-0.432$, which corresponds to log~$R=0$ in Equation~\ref{eqn:orient}.

\noindent At 1450 \AA:
\input{L1450_alpha.tex}
\noindent At 3000 \AA:
\input{L3000_alpha.tex}
\noindent At 5100 \AA:
\input{L5100_alpha.tex}

$\lambda$L$_{\lambda}$ is the uncorrected monochromatic luminosity and $\lambda$L$^{\prime}_{\lambda}$ is the luminosity corrected for $\alpha_{all}$.

The regression and correlation analysis with radio spectral index indicate that a likely orientation dependence is propagated into black hole masses calculated from \Hb\ and \MgII\ FWHM measurements and we again provide a correction for this effect.  We use the fits of $\Delta$log$(M_{BH})($\Hb$-$\CIV$)$ and $\Delta$log$(M_{BH})($\MgII$-$\CIV$)$ to zero the $\alpha_{all}$ dependence, normalizing to $\alpha_{all}=-0.432$.  Using the same methodology as for radio core dominance, we arrive at the following corrections for black hole mass in terms of radio spectral index:

\small
\input{mbhfix_alph.tex}
\normalsize

$M_{BH}(\textrm{\Hb})^{\prime}_{\alpha}$ is the log of the \Hb-based black hole mass with the $\alpha_{all}$ dependence removed, and $M_{BH}(\textrm{\Hb})$ is the log of the orientation-biased, \Hb-based black hole mass estimated from the scaling relationship.  The same conventions apply for \MgII.  The uncertainties given here are the 1-$\sigma$ errors in the fit coefficients and are not independent.  

\begin{figure*}
\begin{minipage}[!b]{8cm}
\centering
\includegraphics[width=8cm]{HbCIVcorrect.eps}
\end{minipage}\hspace{0.6cm}
\hspace{0.6cm}
\begin{minipage}[!b]{8cm}
\centering
\includegraphics[width=8cm]{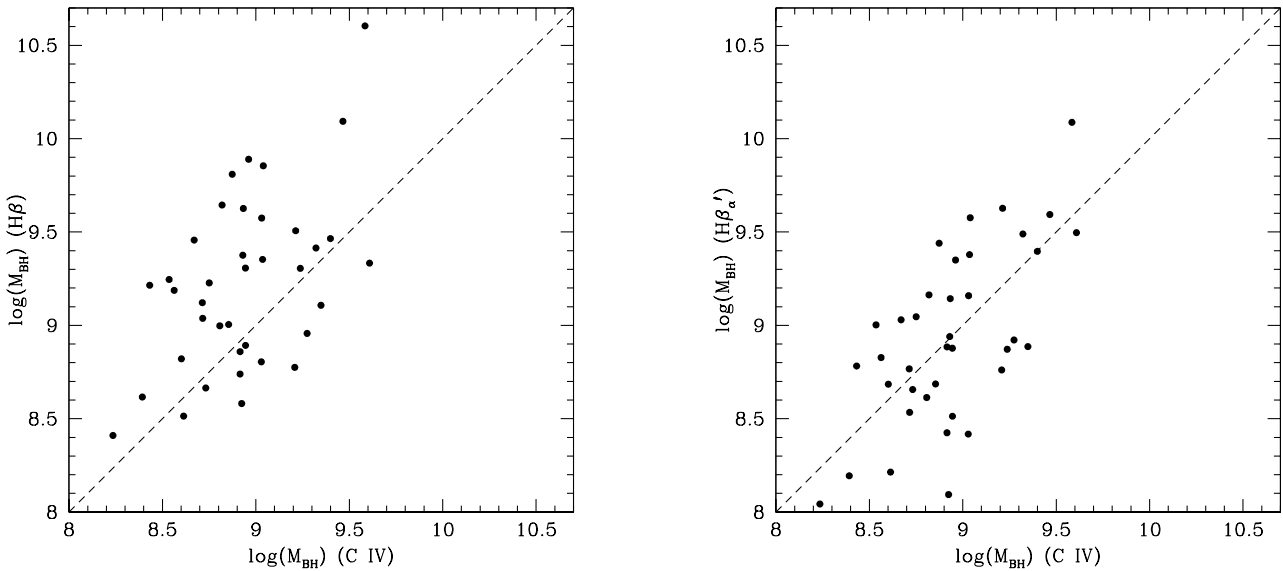}
\end{minipage}
\hspace{0.6cm}                   
\caption{\Hb\ versus \CIV-based black hole masses (left) and radio spectral index corrected \Hb$^{\prime}_{\alpha}$ versus \CIV-based black hole masses (right).  The dashed line is the one-to-one relationship. \label{fig:Hbscatter_alpha}}
\end{figure*}

\begin{figure*}
\begin{minipage}[!b]{8cm}
\centering
\includegraphics[width=8cm]{MgIICIVcorrect.eps}
\end{minipage}\hspace{0.6cm}
\hspace{0.6cm}
\begin{minipage}[!b]{8cm}
\centering
\includegraphics[width=8cm]{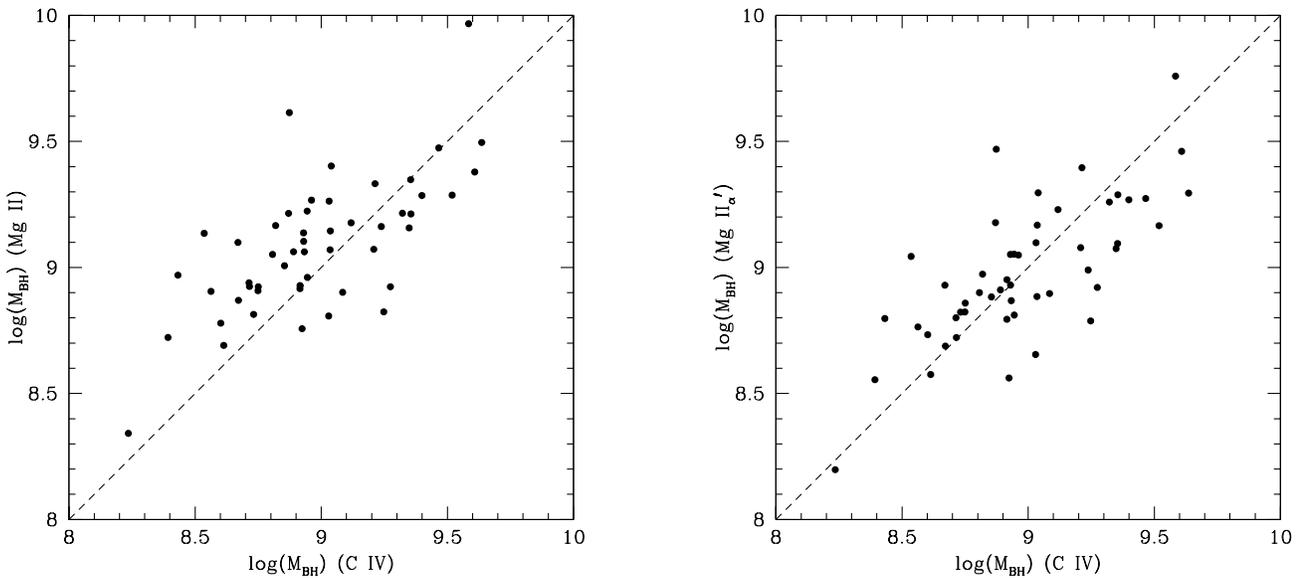}
\end{minipage}
\hspace{0.6cm}                   
\caption{\MgII\ versus \CIV-based black hole masses (left) and radio spectral index corrected \MgII$^{\prime}_{\alpha}$ versus \CIV-based black hole masses (right).  The dashed line is the one-to-one relationship. Note that, while the reduction in scatter is modest, the points now fall more evenly around the one-to-one line.  \label{fig:MgIIscatter_alpha}}
\end{figure*}

Applying this correction reduces the scatter between black hole masses estimated from different broad emission lines.  The scatter around the one-to-one relationship is reduced from $0.40$ dex to $0.35$ dex for \Hb\ and \CIV, $0.194$ dex to $0.188$ dex for \MgII\ and \CIV, and $0.20$ to $0.18$ for \Hb\ and \MgII.  To the number of decimal places shown, the reduction in scatter is exactly the same as for log~$R$.  Figures~\ref{fig:Hbscatter_alpha} and \ref{fig:MgIIscatter_alpha} show the reduction in scatter that results from applying the corrections for $\alpha_{all}$ for \Hb\ and \MgII.  Original masses are shown in the left panel and corrected masses are on the right.

%DISCUSSION
%%%%%%%%%%%%%%%%%%%%%%%%%%%%%%%%%%%%%%%%%%%%%%%%%%%%%%%%%%%%%%%%%%%%%%%%%%%%%%%%%
\section{Discussion}
\label{sec:discussion}
We correlated differences in the log of black hole mass estimates with the orientation indicators radio core dominance and radio spectral index.  We found that the dependence of FWHM of \Hb\ and \MgIIw\ on orientation indicators propagates into black hole masses estimated from scaling relationships based on these emission lines, and derived corrections to \Hb\ and \MgII\ black hole masses for this effect.  The orientation indicator dependence is strongest for the black hole masses based on measurements of \Hb\ FWHM, although differences associated with \MgII-based measurements also show some dependence.

The range in the difference between black hole masses estimated from different emission lines covers nearly an order of magnitude from edge-on to jet-on, although it will depend on the orientation biases in any given sample.  Note that this sample was originally selected to have a full range in log~$R$ and will likely have a different distribution of orientations than a flux limited or randomly selected sample.

Because of the nature of the broad-lined objects in our sample, ``edge-on" will likely be closer to $45^{\circ}$ than $90^{\circ}$ \citep{barthel89}.  If quasars could be observed at a true edge-on inclination, we might expect to measure an even broader FWHM and estimate a correspondingly larger black hole mass.  The scale factor $f$ derived from such a mass might be closer to unity than the larger statistical values that are currently available, implying that the current value is inflated to compensate for orientation and projection effects.

Radio core dominance and radio spectral index are indicators of inclination angle and, while the measurement uncertainties on log~$R$ and $\alpha_{all}$ are relatively low, converting to $\theta$ will introduce additional scatter.  The exact amount is model dependent, but at least in the case of \citet{ghisellini93}, the scatter is substantial.  The implication for this analysis is that the correction to \Hb\ and \MgII\ black hole masses were derived from a fitting procedure that assumed negligible uncertainty in log~$R$ compared to the uncertainty in black hole mass.  A physical correction for inclination angle would require another fitting procedure that assumes uncertainty in both parameters.  We provide only the corrections for log~$R$ and $\alpha_{all}$ as they are model independent and more useful in practice.

Given the size and significance of the R-dependence of black hole mass scaling relationships, it would be advisable to recalibrate them with the radio core dominance dependence taken into account.  The \Hb\ scaling relationship is based on reverberation mapping of the \Hb\ line, so this entails doing reverberation mapping on RL objects and also obtaining UV spectra and assuming that the BLR is the same in RL and radio-quiet objects.  Ideally this would be done for a large sample of RL quasars with known orientations, although few of these have been done.  These new orientation-independent single-epoch mass scaling relationships would be useful for the $\sim 10\%$ of quasars that are RL.  They must also have high-quality radio maps for measuring log~$R$ which becomes more difficult with increasing redshift for a number of reasons.  This could also be done with radio spectral index which is easier to measure and can be measured for some objects in which log~$R$ cannot.

While the \CIV\ scaling relationship does not itself depend on orientation, its original calibration has scatter from orientation effects and should be corrected.  \CIV\ and \MgII\ scaling relationships are based on reverberation mapping of \Hb\ \citep{vestergaard06,vestergaard09} that shows a radio core dominance dependence.  Re-calibrating the \CIV\ and \MgII\ scaling relationships with this effect taken into account should reduce their scatter.

When black hole mass scaling relationships are re-derived, they should also be cast in terms of the line dispersion, $\sigma_{line}$.  The parameter $\sigma_{line}$ is the ``best" line width measure for deriving black hole mass in the sense that it gives the least biased result \citep{peterson11}.  There is a larger range in FWHM than $\sigma_{line}$ and, since the measure of line width is squared in scaling relationships, the extreme values possible for FWHM will introduce a larger range in black hole mass.  If the black hole mass scaling relationships are re-cast in a consistent way for line dispersion as well as FWHM, then the $\sigma_{line}$-based scaling relationships can be explored to see if they are more robust than FWHM against radio core dominance and radio spectral index.

It has been argued that \CIV\ may not be an appropriate choice for estimating black hole mass from a single-epoch spectrum because there may be a strong wind component and the motions of the \CIV\ gas may not be dominated by gravity \citep[e.g.,][]{dunlop04,bachev04,shemmer04,baskin05}.  This is likely the case in many narrow-line Seyfert 1 galaxies that exhibit the characteristic triangular profile indicative of an outflowing wind and the black hole mass scaling relationships do not apply to these objects (none of the objects in this sample have such a line profile for \CIV.)  Quasars do not share these types of asymmetric profiles, but their profiles are often blueshifted relative to the systemic redshift of the object \citep{wilkes84}.  The objects with the largest \CIV\ blueshifts exhibit broader lines \citep{brotherton94,richards02,ho12}, but the corresponding increase in black hole mass is small, well within the uncertainty in the mass estimate.  Despite these issues, it is not clear that \CIV\ is any less trustworthy of a black hole mass indicator than \Hb\ or \MgII, which also have problems \citep{vestergaard11,assef11}.  Reverberation mapping on \CIV\ shows a virial relationship \citep{peterson99,peterson00, vestergaard11, peterson11} and there are a number of advantages to using the \CIV\ line \citep{vestergaard06,warner03,vestergaard02}.  Most importantly for this study, \CIV\ is not biased by orientation as measured by radio core dominance or radio spectral index as are other lines.  The lack of a significant radio core dominance dependence for \CIV\ FWHM may be due to the fact that there is a component to the line that does not reverberate \citep{denney12}, if the reverberating component could be isolated one might expect it to behave more like the \Hb\ line.

The correction for orientation, as indicated by radio core dominance and radio spectral index, is only one of a suite of corrections that will reduce the scatter in black hole mass scaling relationships.  \citet{assef11} perform a study similar to this one on a sample of 12 high-redshift, gravitationally lensed quasars.  They specifically compare black hole masses based on \CIV\ to those from \Hb\ and H$\alpha$ and find that the residuals between the \CIV-based masses and the \Hb\ or H$\alpha$-based masses correlate with the ratio of the UV to optical continuum luminosities.  A correction based on this color dependence reduces the scatter in the residuals by almost a factor of two.  Additional sources of scatter in the black hole mass scaling relationships include the role of radiation pressure in determining the line profile \citep[e.g.,][]{marconi08}, Eddington ratio \citep{onken08}, uncertainties from source variability \citep{woo11}, and effects of low S/N data and measurement uncertainties \citep{denney11}.  While our result is based only on RL quasars, it may be true of RQ quasars as well \citep[as shown by][]{jarvis06}.  If orientation indicators, like that of \citet{boroson11}, are developed for RQ quasars, this effect can be investigated and possibly used to further reduce the scatter in scaling relationships. 

%CONCLUSION
%%%%%%%%%%%%%%%%%%%%%%%%%%%%%%%%%%%%%%%%%%%%%%%%%%%%%%%%%%%%%%%%%%%%%%%%%%%%%%%%%
\section{Conclusions}
\label{sec:conclusion}
Using the RL subsample of the \citet{shang11} SED atlas, we have analyzed the dependence on radio core dominance and radio spectral index for differences in the log of black hole masses estimated from self-consistent scaling relationships and quasi-simultaneous optical-UV data for \Hb, \MgII, and \CIV.  We correlated the difference between the log of black hole masses estimated from these lines with orientation indicators and found that \Hb\ and \MgII-based masses are biased with respect to radio core dominance and radio spectral index as compared to masses estimated from \CIV.  The dependence is such that \Hb\ and \MgII\ will underestimate black hole mass compared to \CIV\ for jet-on objects, the effect is stronger for \Hb, and the range in the difference between black hole masses estimated from different emission lines covers nearly an order of magnitude in black hole mass from our most jet-on to edge-on objects.  We used the \CIV\ FWHM-based mass estimates, unbiased with respect to radio core dominance and radio spectral index, to derive the following corrections for black hole masses based on \Hb\ and \MgII\ FWHM in terms of log~$R$ and $\alpha_{all}$:

\input{mbhfix_conclusion.tex}
\small
\input{mbhfix_conclusion_alph.tex}
\normalsize

When the log~$R$ corrections are applied, scatter around the one-to-one relationship is reduced from 0.40 to 0.35 dex between \Hb\ and \CIV-based black hole masses, 0.194 to 0.188 dex for \MgII\ and \CIV, and 0.20 to 0.18 dex for \Hb\ and \MgII.  The scatter reduction is the same when using corrections based on radio spectral index.

We investigated the role of luminosity in the correlation between differences in the log of black hole mass and orientation indicators and found that differences in luminosity do not depend on radio core dominance or radio spectral index and are therefore not being corrected in our empirical corrections.  The individual luminosities at 1450, 3000, and 5100 \AA\ do depend on the orientation indicators so we provide separate corrections that are to be applied to the luminosity before calculating black hole mass from the single-epoch scaling relationships:\\

\noindent At 1450 \AA:
\input{L1450_conclusion.tex}
\input{L1450_alpha_conclusion.tex}

\noindent At 3000 \AA:
\input{L3000_conclusion.tex}
\input{L3000_alpha_conclusion.tex}
\noindent At 5100 \AA:
\input{L5100_conclusion.tex}
\input{L5100_alpha_conclusion.tex}

This type of correction is only one of a suite of many corrections that will tighten the black hole mass scaling relationships.  By tracking down additional physical parameters that affect mass estimates and searching for new correlations, we can reduce the scatter in the scaling relationships even further.

%ACKNOWLEDGEMENTS
%%%%%%%%%%%%%%%%%%%%%%%%%%%%%%%%%%%%%%%%%%%%%%%%%%%%%%%%%%%%%%%%%%%%%%%%%%%%%%%%%
\section*{Acknowledgments}
J. Runnoe would like to thank Hannah Jang-Condell for her support and Sabrina Cales and Adam Myers for helpful discussions.

This work funded by Wyoming NASA Space Grant Consortium, NASA Grant \#NNX10A095H.
%BIBLIOGRAPHY
%%%%%%%%%%%%%%%%%%%%%%%%%%%%%%%%%%%%%%%%%%%%%%%%%%%%%%%%%%%%%%%%%%%%%%%%%%%%%%%%%
\bibliographystyle{/Users/jrunnoe/Library/texmf/bibtex/bst/mn2e}
\bibliography{./all.092412}
\clearpage

\label{lastpage}
\end{document}

%% file: commands.tex
\newcommand{\CIVw}{C\,{\sc iv}~$\lambda$1549}
\newcommand{\MgIIw}{Mg\,{\sc ii}~$\lambda$2798}

\newcommand{\CIV}{C\,{\sc iv}}
\newcommand{\MgII}{Mg\,{\sc ii}}
\newcommand{\Hb}{H$\beta$}

\newcommand{\FeII}{Fe\,{\sc ii}}
\newcommand{\fwhmhb}{$\textrm{FWHM}_{\textrm{H}\beta}$}
\newcommand{\fwhmciv}{FWHM$_{\textrm{C\,{\sc iv}}}$}
\newcommand{\fwhmmgii}{FWHM$_{\textrm{Mg\,{\sc ii}}}$}
\newcommand{\mbhhb}{M$_{BH}(\textrm{H}\beta)$}
\newcommand{\mbhciv}{M$_{BH}(\textrm{C\,{\sc iv}})$}
\newcommand{\mbhmgii}{M$_{BH}(\textrm{Mg\,{\sc ii}})$}
\def\lsim{\lower0.3em\hbox{$\,\buildrel <\over\sim\,$}}
\def\gsim{\lower0.3em\hbox{$\,\buildrel >\over\sim\,$}}

%% file: radiotab.tex
\begin{table*}
\begin{minipage}[2cm]{14.5cm}
\caption{Radio data \label{tab:radio}}
\renewcommand{\thefootnote}{\alph{footnote}}
\begin{tabular}{lccccrrcrr}
Object & Redshift & $\nu_{core}$/$\nu_{tot}$\footnotemark[1] 	& f$_{core}$ 	   & f$_{ex}$ & log $R$ & log $R^*$ & Reference &$\alpha_{all}$\footnotemark[4] & $\alpha^{1.4}_{4.9}$ \footnotemark[5]\\
       & 	    & (GHz)/(GHz)	      			&	(mJy)      & (mJy)    &         &           &           &               &                      \\
\hline
3C 110   	&      0.7749 &     4.85/    4.85 &        33.38 &        33.63 &   $-$0.25 &         2.68 & 8 & $   -0.99$ & $   -0.65$ \\
3C 175   	&      0.7693 &     8.47/    4.86 &        32.34 &        34.17 &   $-$1.84 &         2.99 & 2,9 & $   -1.02$ & $   -1.10$ \\
3C 186   	&      1.0630 &     1.70/    1.70 &        34.19 &        34.12 & 0.06 &         3.33 & 14 & $   -1.20$ & $   -1.18$ \\
3C 207   	&      0.6797 &     4.86/    1.40 &        33.81 &        34.20 &   $-$0.39 &         3.63 & 2,12 & $   -0.67$ & $   -0.65$ \\
3C 215   	&      0.4108 &     4.89/    4.89 &        31.84 &        33.36 &   $-$1.52 &         3.37 & 6 & $   -0.99$ & $   -1.00$ \\
3C 232   	&      0.5297 &     8.55/    8.55 &        33.64 &        33.00 & 0.64 &         2.87 & 7 & $   -0.51$ & $   -0.41$ \\
3C 254   	&      0.7363 &     5.00/    5.00 &        32.45 &        34.17 &   $-$1.72 &         3.71 & 11 & $   -1.06$ & $   -1.25$ \\
3C 263   	&      0.6464 &     5.00/    4.85 &        33.24 &        34.16 &   $-$0.91 &         3.00 & 2,9 & $   -0.87$ & $   -0.90$ \\
3C 277.1   	&      0.3199 &     1.70/    1.70 &        31.98 &        33.41 &   $-$1.43 &         3.53 & 11 & $   -0.76$ & $   -0.87$ \\
3C 281   	&      0.6017 &     1.40/    1.40 &        32.48 &        33.44 &   $-$0.97 &         3.23 & 10 & $   -0.88$ & $   -0.85$ \\
3C 288.1   	&      0.9631 &     5.00/    5.00 &        31.99 &        34.16 &   $-$2.17 &         3.42 & 11 & $   -0.96$ & $   -1.07$ \\
3C 334   	&      0.5553 &     1.40/    1.40 &        33.18 &        33.83 &   $-$0.65 &         3.11 & 12 & $   -0.87$ & $   -0.83$ \\
3C 37   	&      0.6661 &     5.00/    5.00 &        33.20 &        33.63 &   $-$0.43 &         3.74 & 10 & $   -0.85$ & $   -0.79$ \\
3C 446   	&      1.4040 &     1.40/    1.40 &        35.54 &        34.01\footnotemark[2] & 1.53 &         4.34 & 1 & $   -0.32$ & $   -0.11$ \\
3C 47   	&      0.4250 &     4.89/    4.89 &        32.53 &        33.81 &   $-$1.29 &         3.82 & 6 & $   -0.96$ & $   -0.93$ \\
4C 01.04   	&      0.2634 &     5.00/    5.00 &        32.48 &        33.10 &   $-$0.62 &         3.41 & 13 & $   -0.54$ & $   -0.48$ \\
4C 06.69   	&      1.0002 &     1.40/    1.40 &        34.87 &        32.77\footnotemark[2] & 2.10 &         3.32 & 1 & $    0.16$ & $    0.37$ \\
4C 10.06   	&      0.4075 &     4.99/    4.99 &        32.64 &        33.30 &   $-$0.66 &         2.67 & 18 & $   -0.79$ & $   -0.96$ \\
4C 12.40   	&      0.6836 &     5.00/    5.00 &        31.93 &        33.09\footnotemark[2] &   $-$1.16 &         3.03 & 16 & $   -0.97$ & $   -0.99$ \\
4C 19.44   	&      0.7192 &     1.40/    1.40 &        34.29 &        33.86 & 0.43 &         3.42 & 12 & $   -0.44$ & $   -0.03$ \\
4C 20.24   	&      1.1135 &     1.40/    1.40 &        34.14 &        34.64 &   $-$0.50 &         3.62 & 12 & $   -0.52$ & $   -0.39$ \\
4C 22.26   	&      0.9760 &     1.40/    1.40 &        32.93 &        33.95 &   $-$1.03 &         3.26 & 12 & $   -0.92$ & $   -1.03$ \\
4C 30.25   	&      1.0610 &     1.40/    1.40 &        32.16 &        33.90 &   $-$1.74 &         2.92 & 12 & $   -0.91$ & $   -1.07$ \\
4C 31.63   	&      0.2952 &     1.40/    1.40 &        33.52 &        32.54 & 0.98 &         2.93 & 1 & $   -0.15$ & $    0.00$ \\
4C 39.25   	&      0.6946 &    15.00/   15.00 &        33.46 &        35.68\footnotemark[2] &   $-$2.22 &         3.65 & 3 & $    0.08$ & $    0.75$ \\
4C 40.24   	&      1.2520 &    15.00/   15.00 &        34.60 &        35.08 &   $-$0.48 &         3.94 & 3 & $   -0.19$ & $   -0.01$ \\
4C 41.21   	&      0.6124 &     1.40/    1.40 &        33.40 &        33.88 &   $-$0.48 &         2.91 & 12 & $   -0.79$ & $    0.36$ \\
4C 49.22   	&      0.3333 &     0.40/    0.40 &        32.99 &        33.04 &   $-$0.05 &         3.36 & 3 & $   -0.64$ & $   -0.60$ \\
4C 55.17   	&      0.8990 &     1.40/    1.40 &        34.77 &        33.53 & 1.24 &         3.74 & 12 & $   -0.32$ & $   -0.44$ \\
4C 58.29   	&      1.3740 &     1.40/    1.40 &        32.70 &        34.26 &   $-$1.56 &         2.66 & 12 & $   -0.99$ & $   -1.15$ \\
4C 64.15   	&      1.3000 &     5.00/    5.00 &        32.48 &        34.27 &   $-$1.79 &         3.37 & 11 & $   -0.67$ & $   -0.77$ \\
4C 73.18   	&      0.3027 &     1.40/    1.40 &        33.86 &        32.60 & 1.26 &         3.20 & 1 & $   -0.14$ & $   -0.06$ \\
B2 0742+31   	&      0.4616 &     1.40/    1.40 &        33.52 &        33.16\footnotemark[2] & 0.36 &         2.77 & 12 & $   -0.60$ & $   -0.28$ \\
B2 1351+31   	&      1.3260 &     5.00/    5.00 &        33.22 &        33.70 &   $-$0.47 &         2.95 & 17 & $   -0.79$ & $   -0.80$ \\
B2 1555+33   	&      0.9420 &     1.40/    1.40 &        33.25 &        33.04 & 0.21 &         2.99 & 12 & $   -0.52$ & $   -0.55$ \\
B2 1611+34   	&      1.3945 &    15.00/   15.00 &        35.22 &        35.21 & 0.01 &         3.77 & 3 & $    0.00$ & $   -0.34$ \\
MC2 0042+101   	&      0.5870 &     1.40/    1.40 &        32.15 &        32.98 &   $-$0.82 &         2.92 & 10 & $   -0.87$ & $   -0.78$ \\
MC2 1146+111   	&      0.8614 &     5.00/    5.00 &        31.89 &        33.20 &   $-$1.31 &         2.55 & 16 & $   -0.86$ & $   -0.91$ \\
OS 562   	&      0.7506 &    15.00/   15.00 &        34.41 &        33.52 & 0.89 &         3.35 & 3 & $   -0.08$ & $    0.24$ \\
PG 1100+772   	&      0.3114 &     4.89/    4.89 &        32.23 &        33.32 &   $-$1.10 &         2.65 & 6 & $   -0.84$ & $   -1.00$ \\
PG 1103-006   	&      0.4234 &     1.40/    1.40 &        32.88 &        33.27 &   $-$0.39 &         2.94 & 12 & $   -0.67$ & $   -0.55$ \\
PG 1226+023   	&      0.1576 &     1.40/    1.40 &        34.32 &        33.68 & 0.64 &         3.22 & 12 & $   -0.19$ & $   -0.21$ \\
PG 1545+210   	&      0.2642 &     1.40/    1.40 &        31.50 &        33.25 &   $-$1.75 &         3.00 & 12 & $   -0.74$ & $   -0.77$ \\
PG 1704+608   	&      0.3730 &     1.40/    1.40 &        32.27 &        33.65 &   $-$1.39 &         2.82 & 12 & $   -0.76$ & $   -0.83$ \\
PG 2251+113   	&      0.3253 &     5.00/    5.00 &        30.72 &        33.22 &   $-$2.50 &         2.46 & 15 & $   -0.78$ & $   -0.83$ \\
PKS 0112-017   	&      1.3743 &    15.00/   15.00 &        34.37 &        34.93 &   $-$0.57 &         3.45 & 3 & $    0.13$ & $    0.23$ \\
PKS 0403-13   	&      0.5700 &     1.40/    1.40 &        34.55 &        32.85 & 1.70 &         3.73 & 1 & $   -0.36$ & $   -0.35$ \\
PKS 0859-14   	&      1.3320 &     4.90/    4.90 &        34.94 &        33.54\footnotemark[2] & 1.40 &         3.43 & 8 & $   -0.24$ & $   -0.29$ \\
PKS 1127-14   	&      1.1870 &     1.66/    1.66 &        35.29 &        33.34 & 1.95 &         3.88 & 4 & $    0.04$ & $   -0.21$ \\
PKS 1656+053   	&      0.8890 &     5.00/    5.00 &        33.97 &        33.99\footnotemark[2] &   $-$0.02 &         3.10 & 5 & $   -0.19$ & $   -0.00$ \\
PKS 2216-03   	&      0.8993 &     1.40/    1.40 &        34.57 &        33.65 & 0.92 &         3.23 & 1 & $   -0.08$ & $    0.13$ \\
TEX 1156+213   	&      0.3480 &     1.40/    1.40 &        31.24 &        32.47\footnotemark[2] &   $-$1.23 &         2.38 & 12 & $   -0.75$ & $   -0.79$ \\
\hline
\end{tabular}
\footnotetext[1]{Frequencies are observed-frame.  Fluxes, log~$R$, and $R^{*}$ are k-corrected to 5~GHz and in the rest-frame.}
\footnotetext[2]{Extended radio flux was observed rather than calculated from total and core radio fluxes.}
\footnotetext[3]{{\bf References:} (1) \citet{cooper07}; (2) \citet{mullin06}; (3) \citet{kovalev05}; (4) \citet{stanghellini05}; (5) \citet{scott04}; (6) \citet{gilbert04}; (7) \citet{fey00}; (8) \citet{reid99}; (9) \citet{dennett-thorpe97}; (10) \citet{hutchings96}; (11) \citet{reid95}; (12) \citet{becker95}; (13) \citet{lister94}; (14) \citet{spencer91}; (15) \citet{kellerman89}; (16) \citet{hutchings88}; (17) \citet{rogora86}; (18) \citet{miley78}.}
\footnotetext[4]{Radio spectral index measured from all radio data available in \citet{shang11} SEDs.}
\footnotetext[5]{Radio spectral index measured between observed-frame 1.4 and 4.9 GHz.}
\end{minipage}
\end{table*}

%% file: mbhtab.tex
\begin{table*}
\begin{minipage}[2cm]{17cm}
\caption{Black hole masses and luminosities \label{tab:mbh}}
\begin{tabular}{lccccccc}
Object & log$(L_{bol})$ \footnote{From \citet{runnoe12a}, all other columns from \citet{tang12}.} & \fwhmhb       & \fwhmmgii     & \fwhmciv      & log \mbhhb & log \mbhmgii & log \mbhciv \\
       & (ergs s$^{-1}$) & (km s$^{-1}$) & (km s$^{-1}$) & (km s$^{-1}$) & M$_{\sun}$ &  M$_{\sun}$ &  M$_{\sun}$ \\
\hline
3C 110   	&    46.99 & 12450 &  5645 &  5700 &     10.09 &      9.47 &      9.47 \\
3C 175   	&    46.91 & 20925 &  9515 &  6915 &     10.60 &      9.97 &      9.58 \\
3C 186   	&    46.66 & \nodata &  5790 &  6290 & \nodata &         9.35 &      9.35 \\
3C 207   	&    46.28 &  3505 &  4140 &  4935 &      8.74 &      8.92 &      8.92 \\
3C 215   	&    45.77 &  6760 &  6260 &  5605 &      9.04 &      8.92 &      8.72 \\
3C 232   	&    46.36 &  4655 &  4620 &  7145 &      9.11 &      9.16 &      9.35 \\
3C 254   	&    46.17 & 14095 &  6595 &  5205 &      9.89 &      9.27 &      8.96 \\
3C 263   	&    46.81 &  4970 &  4265 &  3310 &      9.31 &      9.22 &      8.94 \\
3C 277.1   	&    45.61 &  3835 &  3380 &  3215 &      8.41 &      8.34 &      8.24 \\
3C 281   	&    46.23 &  7985 &  5505 &  4865 &      9.38 &      9.10 &      8.93 \\
3C 288.1   	&    46.61 &  8970 &  4320 &  4015 &      9.63 &      9.06 &      8.93 \\
3C 334   	&    46.44 &  6345 &  4840 &  5745 &      9.31 &      9.16 &      9.24 \\
3C 37   	&    46.18 &  4280 &  4200 &  3360 &      8.62 &      8.72 &      8.39 \\
3C 446   	&    47.01 & \nodata &  3955 &  3390 & \nodata &         9.21 &      8.87 \\
3C 47   	&    45.97 & 14005 &  7745 &  5450 &      9.64 &      9.17 &      8.82 \\
4C 01.04   	&    45.44 &  9905 &  9550 &  6665 &      9.25 &      9.14 &      8.54 \\
4C 06.69   	&    47.30 &  4015 &  4220 &  5620 &      9.33 &      9.38 &      9.61 \\
4C 10.06   	&    46.48 &  4735 &  4670 &  3785 &      9.00 &      9.05 &      8.81 \\
4C 12.40   	&    46.17 &  3565 &  4205 &  5300 &      8.58 &      8.76 &      8.92 \\
4C 19.44   	&    46.91 &  4575 &  3105 &  2730 &      9.23 &      8.92 &      8.75 \\
4C 20.24   	&    46.92 & \nodata &  4070 &  3525 & \nodata &         9.14 &      8.93 \\
4C 22.26   	&    46.59 & \nodata &  4910 &  5015 & \nodata &         9.07 &      9.04 \\
4C 30.25   	&    46.27 & \nodata &  4440 &  3730 & \nodata &         8.87 &      8.67 \\
4C 31.63   	&    46.61 &  3395 &  4055 &  4840 &      8.78 &      9.07 &      9.21 \\
4C 39.25   	&    46.91 &  6400 &  5075 &  4775 &      9.51 &      9.33 &      9.21 \\
4C 40.24   	&    46.60 & \nodata &  3650 &  4920 & \nodata &         8.90 &      9.08 \\
4C 41.21   	&    46.75 &  3445 &  3060 &  3800 &      8.80 &      8.81 &      9.03 \\
4C 49.22   	&    45.99 &  3910 &  4415 &  4535 &      8.51 &      8.69 &      8.61 \\
4C 55.17   	&    46.46 & \nodata &  3475 &  6420 & \nodata &         8.82 &      9.25 \\
4C 58.29   	&    47.20 & \nodata &  4830 &  5745 & \nodata &         9.50 &      9.64 \\
4C 64.15   	&    46.73 & \nodata &  5125 &  7245 & \nodata &         9.29 &      9.52 \\
4C 73.18   	&    46.40 &  3095 &  3345 &  3560 &      8.66 &      8.81 &      8.73 \\
B2 0742+31   	&    46.46 & 10690 &  6060 &  4890 &      9.85 &      9.40 &      9.04 \\
B2 1351+31   	&    46.66 & \nodata &  4205 &  3690 & \nodata &         9.06 &      8.89 \\
B2 1555+33   	&    46.21 & \nodata &  4650 &  4240 & \nodata &         8.91 &      8.75 \\
B2 1611+34   	&    47.06 &  4795 &  4010 &  4625 &      9.41 &      9.22 &      9.32 \\
MC2 0042+101   	&    45.68 &  8270 &  6275 &  4195 &      9.21 &      8.97 &      8.43 \\
MC2 1146+111   	&    46.29 &  7835 &  5020 &  3715 &      9.46 &      9.10 &      8.67 \\
OS 562   	&    46.74 &  3305 &  3465 &  3470 &      8.86 &      8.93 &      8.92 \\
PG 1100+772   	&    46.46 &  9390 &  6090 &  4775 &      9.57 &      9.26 &      9.03 \\
PG 1103-006   	&    46.30 &  5270 &  4780 &  4515 &      9.00 &      9.01 &      8.85 \\
PG 1226+023   	&    46.96 &  3405 &  2935 &  4530 &      8.96 &      8.92 &      9.27 \\
PG 1545+210   	&    45.93 &  6885 &  5230 &  4560 &      9.12 &      8.94 &      8.71 \\
PG 1704+608   	&    46.49 & 10465 &  8405 &  4015 &      9.81 &      9.61 &      8.87 \\
PG 2251+113   	&    46.35 &  4060 &  4110 &  4805 &      8.89 &      8.96 &      8.94 \\
PKS 0112-017   	&    46.88 & \nodata &  4130 &  5030 & \nodata &         9.21 &      9.36 \\
PKS 0403-13   	&    46.36 &  3735 &  3535 &  3325 &      8.82 &      8.78 &      8.60 \\
PKS 0859-14   	&    47.22 &  4615 &  3775 &  4520 &      9.46 &      9.29 &      9.40 \\
PKS 1127-14   	&    47.15 & \nodata &  3725 &  3695 & \nodata &         9.18 &      9.12 \\
PKS 1656+053   	&    47.06 &  3510 &  3980 & \nodata &      9.13 &      9.22 & \nodata \\
PKS 2216-03   	&    46.95 &  4415 &  3555 &  3600 &      9.35 &      9.14 &      9.04 \\
TEX 1156+213   	&    46.01 &  7740 &  5175 &  3880 &      9.19 &      8.91 &      8.56 \\
\hline
\end{tabular}
\end{minipage}
\end{table*}

%% file: correlations.tex
\begin{table}
\begin{minipage}[2cm]{8.5cm}
\renewcommand{\thefootnote}{\alph{footnote}}
\caption{Correlations for black hole mass, log~$R$, and $\alpha_{all}$ \label{tab:correlation}}
\begin{tabular}{llcrr}
Param 1 & Param 2 & N & $\rho$ \footnotemark[1]   & P \footnotemark[2]      \\
\hline
 $\Delta$log$(M_{BH})($\Hb$-$\MgII$)$  & log~$R$    & 39 & $  -0.392$ & $   0.014$ \\
 $\Delta$log$(M_{BH})($\Hb$-$\CIV$)$   & log~$R$    & 38 & $  -0.515$ & $   0.001$ \\
 $\Delta$log$(M_{BH})($\MgII$-$\CIV$)$ & log~$R$    & 51 & $  -0.321$ & $   0.022$ \\
 $L_{core}$			    & log~$R$	 & 52 & $   0.837$ & $   <10^{-6}    $ \\
 $L_{ext}$			   	    & log~$R$	 & 52 & $  -0.294$ & $   0.034$ \\
 $\Delta$log$(M_{BH})($\Hb$-$\MgII$)$  & $L_{ext}$  & 39 & $   0.313$ & $   0.052$ \\
 $\Delta$log$(M_{BH})($\Hb$-$\CIV$)$   & $L_{ext}$  & 38 & $   0.228$ & $   0.168$ \\
 $\Delta$log$(M_{BH})($\MgII$-$\CIV$)$ & $L_{ext}$  & 51 & $  -0.154$ & $   0.282$ \\
 $\Delta$log$(M_{BH})($\Hb$-$\MgII$)$  & $L_{core}$ & 39 & $  -0.259$ & $   0.111$ \\
 $\Delta$log$(M_{BH})($\Hb$-$\CIV$)$   & $L_{core}$ & 38 & $  -0.435$ & $   0.006$ \\
 $\Delta$log$(M_{BH})($\MgII$-$\CIV$)$ & $L_{core}$ & 51 & $  -0.449$ & $   0.001$ \\
 $\Delta$log$(M_{BH})($\Hb$-$\MgII$)$  & $\alpha_{all}$    & 39 & $  -0.426$ & $   0.007$ \\
 $\Delta$log$(M_{BH})($\Hb$-$\CIV$)$   & $\alpha_{all}$    & 38 & $  -0.495$ & $   0.002$ \\
 $\Delta$log$(M_{BH})($\MgII$-$\CIV$)$ & $\alpha_{all}$    & 51 & $  -0.340$ & $   0.015$ \\
 log~$R$			    	    & $\alpha_{all}$	& 52 & $   0.625$ & $   <10^{-6}    $ \\
\hline
\end{tabular}
\footnotetext[1]{$\rho$ is the Spearman rank correlation coefficient.}
\footnotetext[2]{P is the given probability that the observed distribution of points would arise by chance.}
\end{minipage}
\end{table}

%% file: L1450.tex
\begin{eqnarray}
\label{eqn:L1450}
\lambda\textrm{L}_{\lambda}^{\prime} = 
\begin{cases} \lambda\textrm{L}_{\lambda}-(   0.292\pm   0.157)\,\textrm{log }R, & \mbox{for log} \,R\geq0 \\
\lambda\textrm{L}_{\lambda}-(   0.097\pm   0.101)\,\textrm{log }R, & \mbox{for log} \,R<0 \end{cases}
\end{eqnarray}

%% file: L3000.tex
\begin{eqnarray}
\label{eqn:L3000}
\lambda\textrm{L}_{\lambda}^{\prime} = 
\begin{cases} \lambda\textrm{L}_{\lambda}-(   0.162\pm   0.135)\,\textrm{log }R, & \mbox{for log} \,R\geq0 \\
\lambda\textrm{L}_{\lambda}-(   0.240\pm   0.104)\,\textrm{log }R, & \mbox{for log} \,R<0 \end{cases}
\end{eqnarray}

%% file: L5100.tex
\begin{eqnarray}
\label{eqn:L5100}
\lambda\textrm{L}_{\lambda}^{\prime} = 
\begin{cases} \lambda\textrm{L}_{\lambda}-(   0.218\pm   0.169)\,\textrm{log }R, & \mbox{for log} \,R\geq0 \\
\lambda\textrm{L}_{\lambda}-(   0.136\pm   0.111)\,\textrm{log }R, & \mbox{for log} \,R<0 \end{cases}
\end{eqnarray}

%% file: mbhfix.tex
\begin{eqnarray}
\label{eqn:Hbnew}
\nonumber \textrm{log }\left[\frac{M_{BH}(\textrm{\Hb})^{\prime}}{M_{\sun}}\right] &=& \textrm{log }\left[\frac{M_{BH}(\textrm{\Hb})}{M_{\sun}}\right] \\
&+& (   0.173\pm   0.051)\,\textrm{log }R 
\end{eqnarray}
\begin{eqnarray}
\label{eqn:MgIInew}
\nonumber \textrm{log }\left[\frac{M_{BH}(\textrm{\MgII})^{\prime}}{M_{\sun}}\right] &=& \textrm{log }\left[\frac{M_{BH}(\textrm{\MgII})}{M_{\sun}}\right] \\
 &+& (   0.066\pm   0.028)\,\textrm{log }R
\end{eqnarray}

%% file: orient.tex
\begin{eqnarray}
\label{eqn:orient}
\alpha_{all} &=& (   0.260\pm   0.048)\,\textrm{log }R - (   0.433\pm   0.048)
\end{eqnarray}

%% file: L1450_alpha.tex
\begin{eqnarray}
\label{eqn:L1450_alph}
\lambda\textrm{L}_{\lambda}^{\prime} &=& \lambda\textrm{L}_{\lambda} - (   0.716\pm   0.202)\,\alpha_{all} - (   0.310\pm   0.211)
\end{eqnarray}

%% file: L3000_alpha.tex
\begin{eqnarray}
\label{eqn:L3000_alph}
\lambda\textrm{L}_{\lambda}^{\prime} &=& \lambda\textrm{L}_{\lambda} - (   0.678\pm   0.201)\,\alpha_{all} - (   0.293\pm   0.210)
\end{eqnarray}

%% file: L5100_alpha.tex
\begin{eqnarray}
\label{eqn:L5100_alph}
\lambda\textrm{L}_{\lambda}^{\prime} &=& \lambda\textrm{L}_{\lambda} - (   0.675\pm   0.222)\,\alpha_{all} - (   0.292\pm   0.229)
\end{eqnarray}

%% file: mbhfix_alph.tex
\begin{eqnarray}
\label{eqn:Hbnew_alph}
\nonumber \textrm{log }\left[\frac{M_{BH}(\textrm{\Hb})^{\prime}_{\alpha}}{M_{\sun}}\right] &=& \textrm{log }\left[\frac{M_{BH}(\textrm{\Hb})}{M_{\sun}}\right] + (   0.580\pm   0.170)\,\alpha_{all} \\
&+& (   0.251\pm   0.180)
\end{eqnarray}
\begin{eqnarray}
\label{eqn:MgIInew_alph}
\nonumber \textrm{log }\left[\frac{M_{BH}(\textrm{\MgII})^{\prime}_{\alpha}}{M_{\sun}}\right] &=& \textrm{log }\left[\frac{M_{BH}(\textrm{\MgII})}{M_{\sun}}\right] + (   0.247\pm   0.088)\,\alpha_{all} \\
&+& (   0.107\pm   0.106)
\end{eqnarray}

%% file: mbhfix_conclusion.tex
\begin{eqnarray}
\nonumber \textrm{log }\left[\frac{M_{BH}(\textrm{\Hb})^{\prime}}{M_{\sun}}\right] &=& \textrm{log }\left[\frac{M_{BH}(\textrm{\Hb})}{M_{\sun}}\right] \\
&+& (   0.173\pm   0.051)\,\textrm{log }R 
\end{eqnarray}
\begin{eqnarray}
\nonumber \textrm{log }\left[\frac{M_{BH}(\textrm{\MgII})^{\prime}}{M_{\sun}}\right] &=& \textrm{log }\left[\frac{M_{BH}(\textrm{\MgII})}{M_{\sun}}\right] \\
 &+& (   0.066\pm   0.028)\,\textrm{log }R
\end{eqnarray}

%% file: mbhfix_conclusion_alph.tex
\begin{eqnarray}
\nonumber \textrm{log }\left[\frac{M_{BH}(\textrm{\Hb})^{\prime}_{\alpha}}{M_{\sun}}\right] &=& \textrm{log }\left[\frac{M_{BH}(\textrm{\Hb})}{M_{\sun}}\right] + (   0.580\pm   0.170)\,\alpha_{all} \\
&+& (   0.251\pm   0.180)
\end{eqnarray}
\begin{eqnarray}
\nonumber \textrm{log }\left[\frac{M_{BH}(\textrm{\MgII})^{\prime}_{\alpha}}{M_{\sun}}\right] &=& \textrm{log }\left[\frac{M_{BH}(\textrm{\MgII})}{M_{\sun}}\right] + (   0.247\pm   0.088)\,\alpha_{all} \\
&+& (   0.107\pm   0.106)
\end{eqnarray}

%% file: L1450_conclusion.tex
\begin{eqnarray}
\lambda\textrm{L}_{\lambda}^{\prime} = 
\begin{cases} \lambda\textrm{L}_{\lambda}-(   0.292\pm   0.157)\,\textrm{log }R, & \mbox{for log} \,R\geq0 \\
\lambda\textrm{L}_{\lambda}-(   0.097\pm   0.101)\,\textrm{log }R, & \mbox{for log} \,R<0 \end{cases}
\end{eqnarray}

%% file: L1450_alpha_conclusion.tex
\begin{eqnarray}
\lambda\textrm{L}_{\lambda}^{\prime} &=& \lambda\textrm{L}_{\lambda} - (   0.716\pm   0.202)\,\alpha_{all} - (   0.310\pm   0.211)
\end{eqnarray}

%% file: L3000_conclusion.tex
\begin{eqnarray}
\lambda\textrm{L}_{\lambda}^{\prime} = 
\begin{cases} \lambda\textrm{L}_{\lambda}-(   0.162\pm   0.135)\,\textrm{log }R, & \mbox{for log} \,R\geq0 \\
\lambda\textrm{L}_{\lambda}-(   0.240\pm   0.104)\,\textrm{log }R, & \mbox{for log} \,R<0 \end{cases}
\end{eqnarray}

%% file: L3000_alpha_conclusion.tex
\begin{eqnarray}
\lambda\textrm{L}_{\lambda}^{\prime} &=& \lambda\textrm{L}_{\lambda} - (   0.678\pm   0.201)\,\alpha_{all} - (   0.293\pm   0.210)
\end{eqnarray}

%% file: L5100_conclusion.tex
\begin{eqnarray}
\lambda\textrm{L}_{\lambda}^{\prime} = 
\begin{cases} \lambda\textrm{L}_{\lambda}-(   0.218\pm   0.169)\,\textrm{log }R, & \mbox{for log} \,R\geq0 \\
\lambda\textrm{L}_{\lambda}-(   0.136\pm   0.111)\,\textrm{log }R, & \mbox{for log} \,R<0 \end{cases}
\end{eqnarray}

%% file: L5100_alpha_conclusion.tex
\begin{eqnarray}
\lambda\textrm{L}_{\lambda}^{\prime} &=& \lambda\textrm{L}_{\lambda} - (   0.675\pm   0.222)\,\alpha_{all} - (   0.292\pm   0.229)
\end{eqnarray}